\documentclass[11pt,onecolumn]{IEEEtran}

\usepackage{color}
\usepackage{amsmath}
\usepackage{amssymb}
\usepackage[dvips]{graphicx}
\usepackage{setspace}

\renewcommand{\baselinestretch}{1.7}

\newcommand{\mm}{{\bf{m}}}

\newcommand{\llv}{{\boldsymbol{l}}}
\newcommand{\zz}{{\bf{z}}}
\newcommand{\ff}{{\bf{f}}}

\newcommand{\bet}{{\boldsymbol{\eta}}}
\newcommand{\blam}{{\boldsymbol{\lambda}}}
\newcommand{\bthe}{{\boldsymbol{\theta}}}

\newcommand{\rmE}{{\rm{E}}}

\newcommand{\Tf}{T_{\rm{f}}}

\newcommand{\Nf}{N_{\rm{f}}}
\newcommand{\Pmax}{P_{\rm{max}}}

\begin{document}

\title{Position Estimation via Ultra-Wideband Signals
\thanks{This work was supported in part by the European Commission in
the framework of the FP7 Network of Excellence in Wireless
COMmunications NEWCOM++ (contract n. 216715), and in part by the U.
S. National Science Foundation under Grants ANI-03-38807 and
CNS-06-25637.}}

\author{Sinan Gezici\thanks{Sinan Gezici is with the Department of
Electrical and Electronics Engineering, Bilkent University, Bilkent,
Ankara TR-06800, Turkey, Tel: +90 (312) 290-3139, Fax: +90 (312)
266-4192, e-mail: gezici@ee.bilkent.edu.tr.}, \textit{Member, IEEE,}
and H. Vincent Poor\thanks{H. Vincent Poor is with the Department of
Electrical Engineering, Princeton University, Princeton 08544, USA,
Tel: (609) 258-2260, Fax: (609) 258-7305, email:
poor@princeton.edu.}, \textit{Fellow, IEEE}}

\maketitle

\begin{abstract}
The high time resolution of ultra-wideband (UWB) signals facilitates
very precise position estimation in many scenarios, which makes a variety
applications possible. This paper reviews the problem of
position estimation in UWB systems, beginning with an overview of
the basic structure of UWB signals and their positioning
applications.  This overview is followed by a discussion of  various
position estimation techniques, with an emphasis on time-based
approaches, which are particularly suitable for UWB positioning
systems.  Practical issues arising in UWB signal design and hardware
implementation are also discussed.
\end{abstract}


\section{Ultra-Wideband Signals and Positioning Applications}\label{sec:UWB_apps}

\subsection{Ultra-Wideband Signals}

\begin{figure}
\begin{center}
\includegraphics[width=0.65\textwidth]{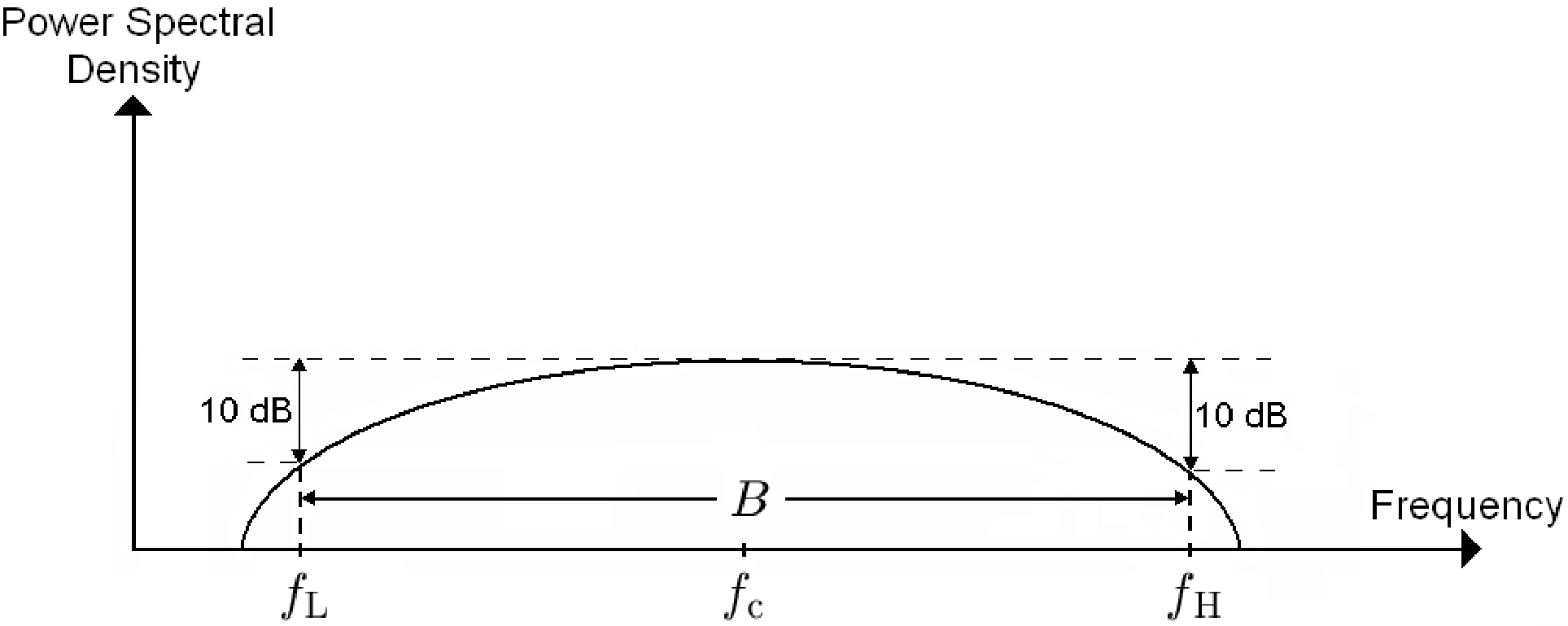}
\caption{A UWB signal is defined to have an
absolute bandwidth $B$ of at least $500$ MHz, or a fractional
bandwidth $B_{\rm{frac}}$ larger than $0.2$ (c.f.
(\ref{eq:fracBW2})) \cite{ourBook}.} \label{fig:spectrum}
\end{center}
\end{figure}

Ultra-wideband (UWB) signals are characterized by their very large
bandwidths compared to those of conventional narrowband/wideband
signals. According to the U.S. Federal Communications Commission
(FCC), a UWB signal is defined to have an absolute bandwidth of at
least $500$ MHz or a fractional (relative) bandwidth of larger than
20\% \cite{FCC_0248}. As shown in Fig. \ref{fig:spectrum}, the
absolute bandwidth is obtained as the difference between the upper
frequency $f_{\rm{H}}$ of the $-10\,\rm{dB}$ emission point and the
lower frequency $f_{\rm{L}}$ of the $-10\,\rm{dB}$ emission point;
i.e.,
\begin{gather}\label{eq:absBW}
B=f_{\rm{H}}-f_{\rm{L}}~,
\end{gather}
which is also called \textit{$-10$ dB bandwidth}. On the other hand,
the fractional bandwidth is calculated as
\begin{gather}\label{eq:fracBW}
B_{\rm{frac}}=\frac{B}{f_{\rm{c}}}~,
\end{gather}
where $f_c$ is the center frequency and is given by
\begin{gather}\label{eq:centerFreq}
f_c=\frac{f_{\rm{H}}+f_{\rm{L}}}{2}~.
\end{gather}
From (\ref{eq:absBW}) and (\ref{eq:centerFreq}), the fractional
bandwidth $B_{\rm{frac}}$ in (\ref{eq:fracBW}) can also be expressed
as
\begin{gather}\label{eq:fracBW2}
B_{\rm{frac}}=\frac{2(f_{\rm{H}}-f_{\rm{L}})}{f_{\rm{H}}+f_{\rm{L}}}~.
\end{gather}

\begin{figure}
\begin{center}
\includegraphics[width=0.65\textwidth]{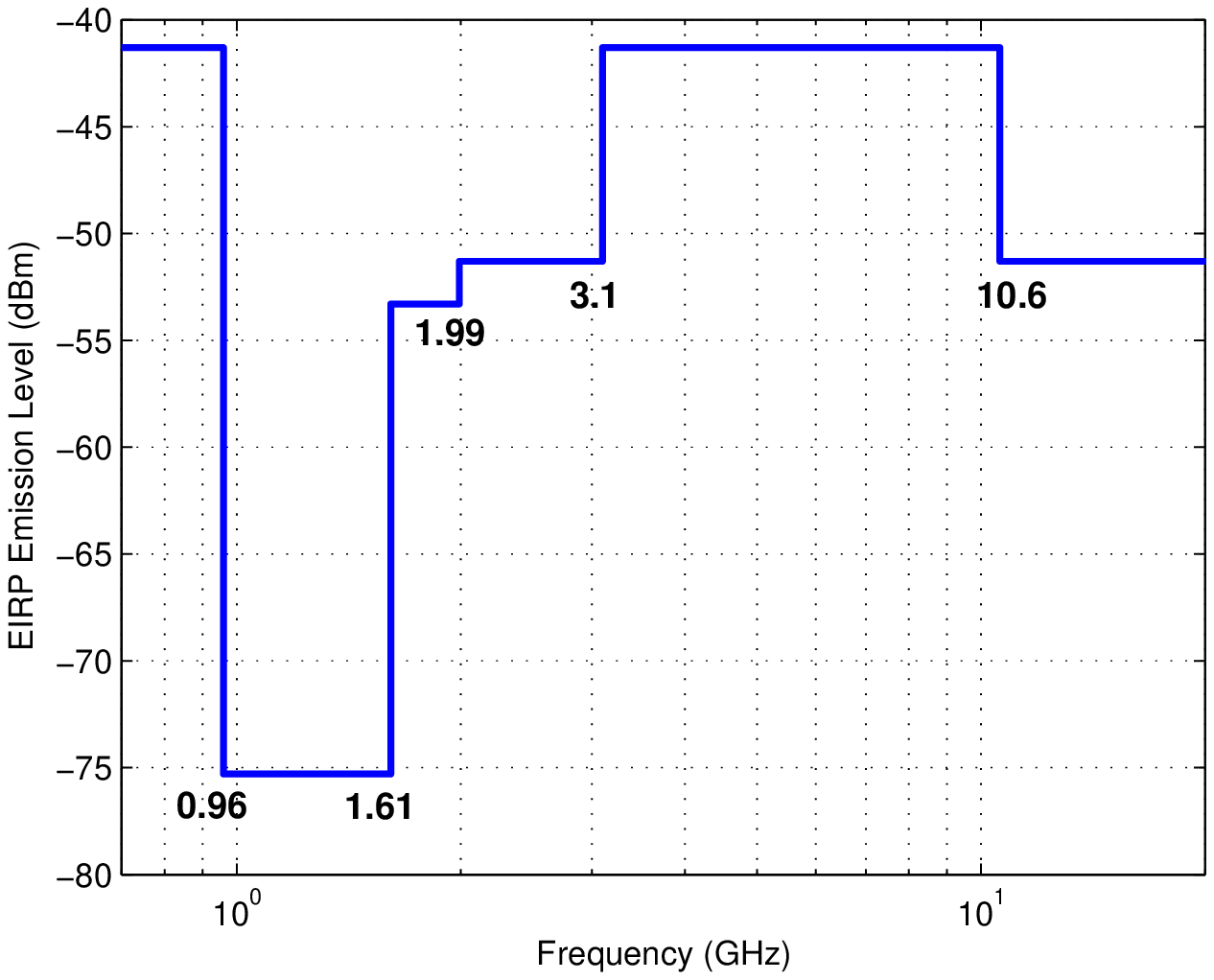}
\caption{FCC emission limits for indoor UWB systems. Please refer to
\cite{FCC_0248} for the regulations for imaging, vehicular radar,
and outdoor communications systems. Note that the limits are
specified in terms of equivalent isotropically-radiated power
(EIRP), which is defined as the product of the power supplied to an
antenna and its gain in a given direction relative to an isotropic
antenna. According to the FCC regulations, emissions (EIRPs) are
measured using a resolution bandwidth of $1$ MHz.}
\label{fig:indoor}
\end{center}
\end{figure}

As UWB signals occupy a very large portion in the spectrum, they
need to coexist with the incumbent systems without causing
significant interference. Therefore, a set of regulations are
imposed on systems transmitting UWB signals. According to the FCC
regulations, UWB systems must transmit below certain power levels in
order not to cause significant interference to the legacy systems in
the same frequency spectrum. Specifically, the average power
spectral density (PSD) must not exceed $-41.3$ dBm/MHz over the
frequency band from $3.1$ to $10.6$ GHz, and it must be even lower
outside this band, depending on the specific application
\cite{FCC_0248}. For example, Fig. \ref{fig:indoor} illustrates the
FCC limits for indoor communications systems.
After the FCC legalized the use of UWB signals in the U.S.,
considerable amount of effort has been put into development and
standardization of UWB systems \cite{IEEE802154aD4}, \cite{ecma368}.
Also, both Japan and Europe have recently allowed the use of UWB
systems under certain regulations \cite{Kohno_IEEE}, \cite{ECC_new}.

\begin{figure}
\begin{center}
\includegraphics[width=0.65\textwidth]{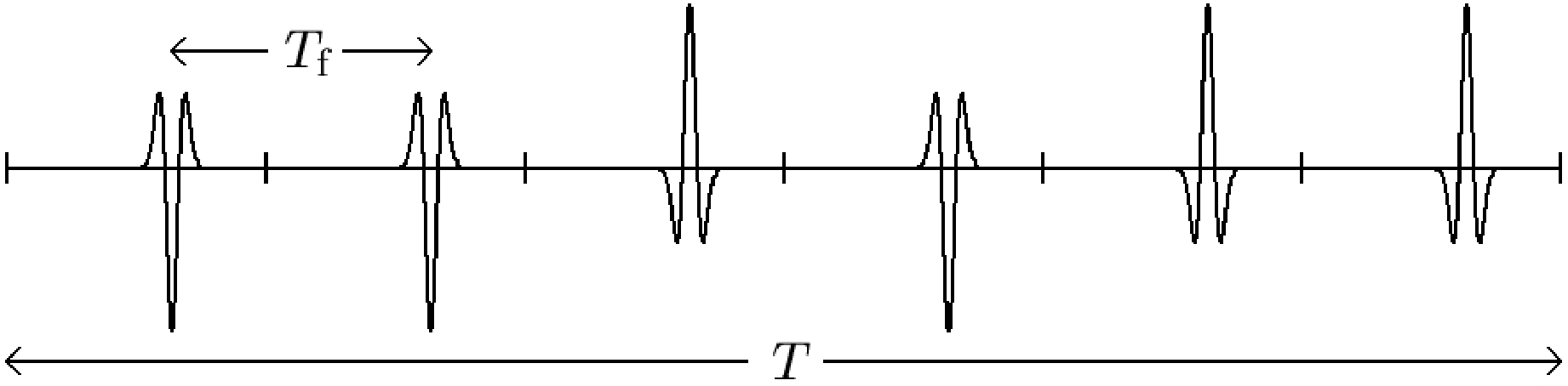}
\caption{An example UWB signal consisting of short duration pulses
with a low duty cycle, where $T$ is the signal duration, and $\Tf$
represents the pulse repetition interval or the frame interval.}
\label{fig:signal}
\end{center}
\end{figure}

Because of the inverse relation between the bandwidth and the
duration of a signal, UWB systems are characterized by very short
duration waveforms, usually on the order of a nanosecond. Commonly,
a UWB system transmits very short duration pulses with a low duty
cycle; that is, the ratio between the pulse transmission instant and
the average time between two consecutive transmissions is usually
kept small, as shown in Fig. \ref{fig:signal}.
Such a pulse-based UWB signaling scheme is called \textit{impulse
radio (IR)} UWB \cite{scholtz}. In an IR UWB communications system,
a number of UWB pulses are transmitted per information symbol and
information is usually conveyed by the timings or
the polarities of the pulses\footnote{In addition to IR UWB systems,
it is also possible to realize UWB systems with continuous transmissions. For example, direct sequence code division multiple access (DS-CDMA) systems with very short chip intervals can be classified as a UWB communications system \cite{DS_UWB}. Alternatively, transmission and reception of very short duration
orthogonal frequency division multiplexing (OFDM) symbols can be considered as an OFDM UWB scheme \cite{MB_OFDM_UWB}. However, the focus of this paper will be on IR UWB systems.}. For positioning systems, the main purpose is to estimate
position related parameters of this IR UWB signal, such as its time-of-arrival (TOA), as will be discussed in Section \ref{sec:posEst}.

Large bandwidths of UWB signals bring many advantages for
positioning, communications, and radar applications \cite{ourBook}:
\begin{itemize}
\item Penetration through obstacles \item Accurate position estimation
\item High-speed data transmission \item Low
cost and low power transceiver designs
\end{itemize}

The penetration capability of a UWB signal is due to its large
frequency spectrum that includes low frequency components as well as
high frequency ones. This large spectrum also results in high time
resolution, which improves ranging (i.e., distance estimation)
accuracy, as will be discussed in Section \ref{sec:posEst}.

The appropriateness of UWB signals for high-speed data
communications can be observed from the Shannon capacity formula.
For an additive white Gaussian noise (AWGN) channel with bandwidth
of $B$ Hz, the maximum data rate that can be transmitted to a
receiver with negligible error is given by
\begin{gather}\label{eq:capacity}
C=B\log_2(1+\rm{SNR})\quad (\rm{bits/second}),
\end{gather}
where $\rm{SNR}$ is the signal-to-noise ratio of the system. In
other words, as the bandwidth of the system increases, more
information can be sent from the transmitter to the receiver. Also
note that for large bandwidths, signal power can be kept at low
levels in order to increase the battery life of the system and to
minimize the interference to the other systems in the same frequency
spectrum.

Moreover, a UWB system can be realized in baseband (carrier-free),
that is UWB pulses can be transmitted without a sine-wave carrier.
In that case, it becomes possible to design transmitters and
receivers with fewer components \cite{ourBook}.

\subsection{UWB Positioning Applications}

For positioning systems, UWB signals provide an accurate, low cost
and low power solution thanks to their unique properties discussed
above. Especially, short-range wireless sensor networks (WSNs),
which combine low/medium data rate communications with positioning
capability seem to be the emerging application of UWB signals
\cite{Gezici_SPM_2005}. Some important applications of UWB WSNs can
be exemplified as follows \cite{ourBook}, \cite{Gezici_SPM_2005},
\cite{CallForAppl_Siwiak}:
\begin{itemize}
\item \textbf{Medical:} Wireless body area networking for
fitness and medical purposes, and monitoring the locations of
wandering patients in an hospital. \item \textbf{Security/Military:}
Locating authorized people in high-security areas and tracking the
positions of the military personnel. \item \textbf{Inventory
Control:} Real-time tracking of shipments and valuable items in
manufacturing plants, and locating medical equipments in hospitals.
\item \textbf{Search and Rescue:} Locating lost children, injured sportsmen,
emergency responders, miners, avalanche/earthquake victims, and
fire-fighters. \item \textbf{Smart Homes: } Home security, control
of home appliances, and locating inhabitants.
\end{itemize}

Accuracy requirements of these positioning scenarios vary depending
on the specific application \cite{CallForAppl_Siwiak}. For most
applications, an accuracy of less than a foot is
desirable, which makes UWB signaling a unique candidate in those
scenarios.

The opportunities offered by UWB WSNs also resulted in the formation
of the IEEE 802.15 low rate alternative PHY task group (TG4a) in
2004 to design an alternate PHY specification for the already
existing IEEE 802.15.4 standard for wireless personal area networks
(WPANs) \cite{OLD_IEEE802154}. The main aim of the TG4a was to
provide communications and high-precision positioning with low power
and low cost devices \cite{IEEE_4a}. In March 2007, IEEE 802.15.4a
was approved as a new amendment to IEEE Std 802.15.4-2006. The 15.4a
amendment specifies two optional signaling formats based on UWB and
chirp spread spectrum (CSS) signaling \cite{IEEE802154aD4}. The UWB
option can use $250-750$ MHz, $3.244-4.742$ GHz, or $5.944-10.234$
GHz bands; whereas the CSS uses the $2.4-2.4835$ GHz band. Although
the CSS option can only be used for communications purposes, the UWB
option has an optional ranging capability, which facilitates new
applications and market opportunities offered by UWB positioning
systems.

UWB positioning systems have also attracted
significant interest from the research community. Recent books on UWB
systems and in general on wireless networks study UWB positioning
applications as well
\cite{BookOpperman}-\nocite{BookPahlavan}\cite{BookKohno}. In
addition, research articles on UWB positioning, such as
\cite{Gezici_SPM_2005} and references therein, consider various
aspects of position estimation based on UWB signals. The main
purpose of this article is to present a general overview of UWB
positioning systems and present not only signal processing issues as
in \cite{Gezici_SPM_2005} but also practical design constraints,
such as limitations on hardware components.


\section{Position Estimation Techniques}\label{sec:posEst}

\begin{figure}
\begin{center}
\includegraphics[width=0.65\textwidth]{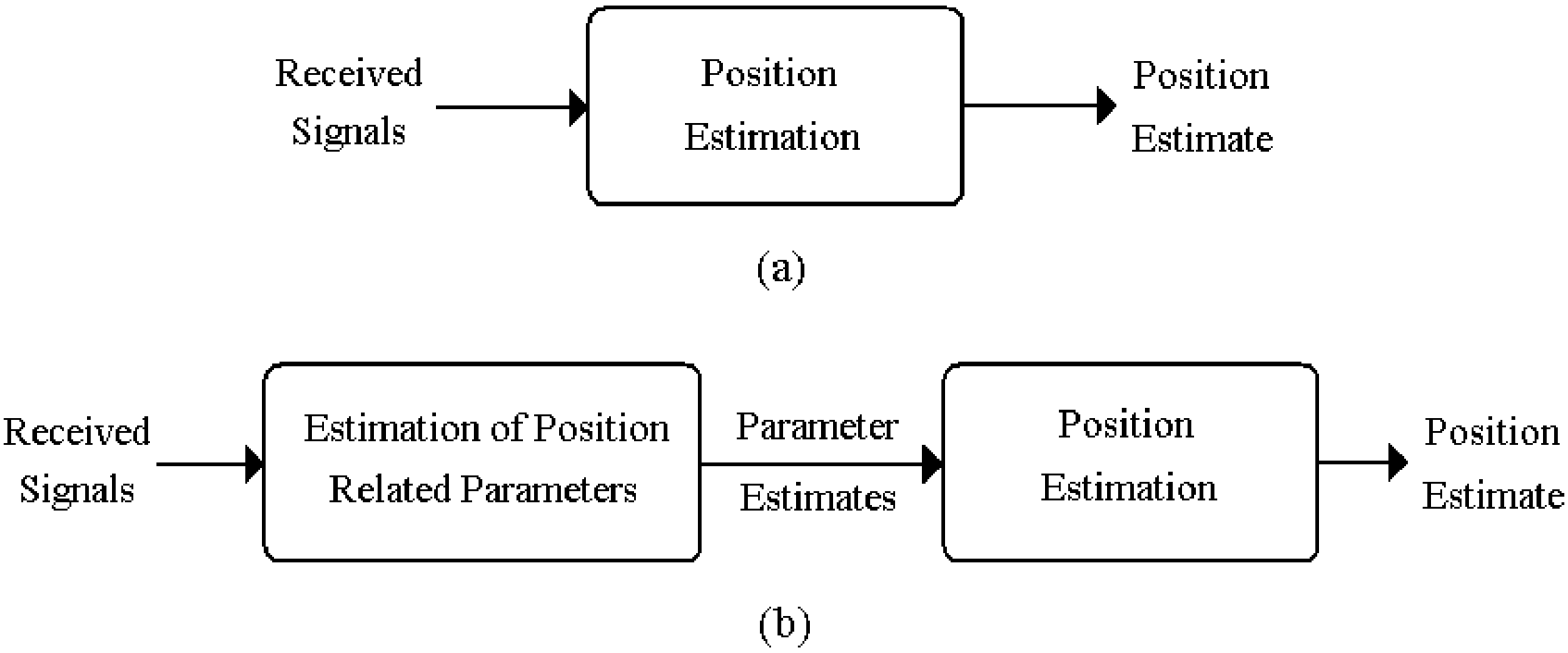}
\caption{(a) Direct positioning, (b) two-step positioning \cite{Sinan_Springer07}.}
\label{fig:direct_OR_2step}
\end{center}
\end{figure}

In order to comprehend the high-precision positioning capability of
UWB signals, position estimation techniques should be investigated
first. Position estimation of a node\footnote{A ``node'' refers to
any device involved in the position estimation process, such as a
wireless sensor or a base station.} in a wireless network involves
signal exchanges between that node (called the ``target'' node;
i.e., the node to be located) and a number of reference
nodes \cite{Sinan_Springer07}. The position of
a target node can be estimated by the target node itself, which is
called \textit{self-positioning}, or it can be estimated by a
central unit that gathers position information from the reference
nodes, which is called \textit{remote-positioning}
(\textit{network-centric positioning}) \cite{Gustafsson_SPM_2005}.
In addition, depending on whether the position is estimated from the
signals traveling between the nodes directly or not, two different
position estimation schemes can be considered, as shown in Fig.
\ref{fig:direct_OR_2step} \cite{ourBook}, \cite{Sinan_Springer07}.
\textit{Direct positioning} refers to the case in which the position
is estimated directly from the signals traveling between the nodes
\cite{Weiss_SPL_2004}. On the other hand, a \textit{two-step
positioning} system first extracts certain signal parameters from
the signals, and then estimates the position based on those signal
parameters. Although the two-step positioning approach is suboptimal
in general, it can have significantly lower complexity than the
direct approach. Also, the performance of the two approaches is
usually quite close for sufficiently high signal bandwidths and/or
signal-to-noise ratios (SNRs) \cite{Weiss_SPL_2004},
\cite{Qi_TWC_2006}. Therefore, the two-step positioning is the
common technique in most positioning systems, which will also be the
main focus of this paper.

In the first step of a two-step positioning technique, signal
parameters, such as time-of-arrival (TOA), angle-of-arrival (AOA),
and/or received signal strength (RSS), are estimated. Then, in the
second step, the target node position is estimated based on the
signal parameters obtained from the first step (Fig.
\ref{fig:direct_OR_2step}-(b)). In the following, various techniques
for this two-step positioning approach are studied in detail.

\subsection{Estimation of Position Related Parameters}\label{sec:parEst}

As shown in Fig. \ref{fig:direct_OR_2step}-(b), the first step in a
two-step positioning algorithm involves the estimation of parameters
related to the position of the target node. Those parameters are
usually related to the energy, timing and/or direction of the
signals traveling between the target node and the reference nodes.
Although it is common to estimate a single parameter for each signal
between the target node and a reference node, such as the arrival
time of the signal, it is also possible to estimate multiple
position related parameters per signal in order to improve
positioning accuracy.

\subsubsection{Received Signal Strength}\label{sec:RSS}

As the energy of a signal changes with distance, the received signal
strength (RSS) at a node conveys information about the distance
(``range'') between that node and the node that has transmitted the
signal. In order to convert the RSS information into a range
estimate, the relation between distance and signal energy should be
known. In the presence of such a relation, the distance between the
nodes can be estimated from the RSS measurement at one of the nodes
assuming that the transmitted signal energy is known.

One factor that affects the signal energy is called
\textit{path-loss}, which refers to the reduction of signal
power/energy as it propagates through space. A common model for
path-loss is given by
\begin{gather}\label{eq:avgPower}
\bar{P}(d)=P_0-10n\log_{10}(d/d_0),
\end{gather}
where $n$ is called the \textit{path-loss exponent}, $\bar{P}(d)$ is
the average received power in dB at a distance $d$, and $P_0$ is the
received power in dB at a short reference distance $d_0$. The
relation in (\ref{eq:avgPower}) specifies the relation between the
power loss and distance through the path-loss exponent.

Although there is a simple relation between \textit{average} signal
power and distance as shown in (\ref{eq:avgPower}), the exact
relation between distance and signal energy in a practical wireless
environment is quite complicated due to propagation mechanisms such
as reflection, scattering, and diffraction, which can cause
significant fluctuations in RSS even over short distances and/or
small time intervals. In order to obtain a reliable range estimate,
signal power is commonly obtained as
\begin{gather}\label{eq:recPower}
P(d)=\frac{1}{T}\int_{0}^{T}|r(t,d)|^2dt,
\end{gather}
where $r(t,d)$ is the received signal at distance
$d$ and $T$ is the integration interval. Although the averaging
operation in (\ref{eq:recPower}) can mitigate the short-term
fluctuations called \textit{small-scale fading}, the average power
(or RSS) still varies about its local mean, given by
(\ref{eq:avgPower}), due to \textit{shadowing} effects, which
represent signal energy variations due to the obstacles in the
environment. Shadowing is commonly modeled by a zero-mean Gaussian
random variable in the logarithmic scale. Therefore, the received
power $P(d)$ in dB can be expressed as\footnote{There is also
thermal noise in practical systems, which is commonly
location-dependent. In this study, it is assumed that the thermal
noise is sufficiently mitigated \cite{Yihong_tez}.}
\begin{gather}\label{eq:RSSmodel}
P(d)\sim{\mathcal{N}}\left(\bar{P}(d)\,,\,\sigma^2_{\rm{sh}}\right),
\end{gather}
where $\bar{P}(d)$ is as given in (\ref{eq:avgPower}), and
$\sigma^2_{\rm{sh}}$ is the variance of the log-normal shadowing
variable.

From (\ref{eq:RSSmodel}), it is observed that accurate knowledge of
the path-loss exponent and the shadowing variance is required for a
reliable range estimate based on RSS measurements. How accurate a
range estimate can be obtained is specified by a lower bound, called
the Cramer-Rao lower bound (CRLB), on the variance of an
unbiased\footnote{For an unbiased estimate, the mean (expected
value) of the estimate is equal to the true value of the parameter
to be estimated.} range estimate \cite{Yihong_tez};
\begin{gather}\label{eq:CRLB_RSS}
\sqrt{{\rm{Var}}\{\hat{d}\}}\geq
\frac{(\ln10)\,\sigma_{\rm{sh}}\,d}{10\,n}\,,
\end{gather}
where $\hat{d}$ represents an unbiased estimate for the distance
$d$. Note from (\ref{eq:CRLB_RSS}) that the range estimates get more
accurate as the standard deviation of the shadowing decreases, which
makes RSS vary less around the true average power. Also, a larger
path-loss exponent results in a smaller lower bound, since the
average power becomes more sensitive to distance for larger $n$.
Finally, the accuracy of the range estimates deteriorates as the
distance between the nodes increases.

Commonly, the RSS technique cannot provide very accurate range
estimates due to its heavy dependence on the channel parameters,
which is also true for UWB systems. For example, in a
non-line-of-sight (NLOS) residential environment, modeled according
to the IEEE 802.15.4a UWB channel model \cite{IEEEChanMod4a}, with
$n=4.58$ and $\sigma_{\rm{sh}}=3.51$, the lower bound in
(\ref{eq:CRLB_RSS}) is about $1.76$ m. at $d=10$ m.

\subsubsection{Angle of Arrival}\label{sec:AOA}

\begin{figure}
\begin{center}
\includegraphics[width=0.15\textwidth]{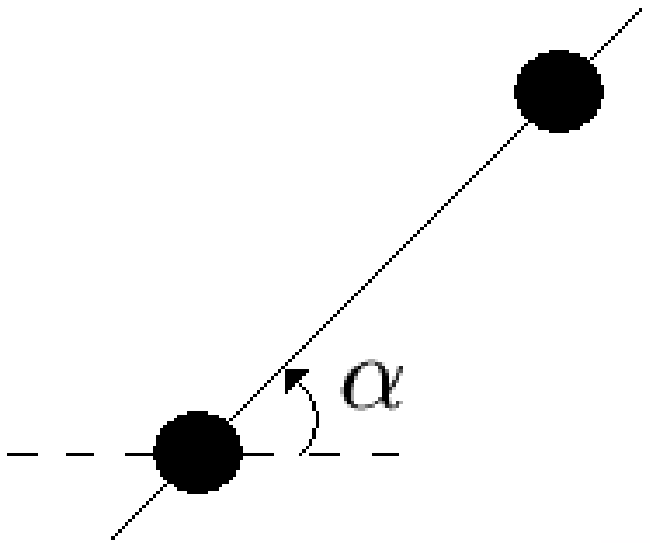}
\caption{The AOA measurement at a node gives information about the
direction over which the target node lies.} \label{fig:AOA}
\end{center}
\end{figure}

\begin{figure}
\begin{center}
\includegraphics[width=0.175\textwidth]{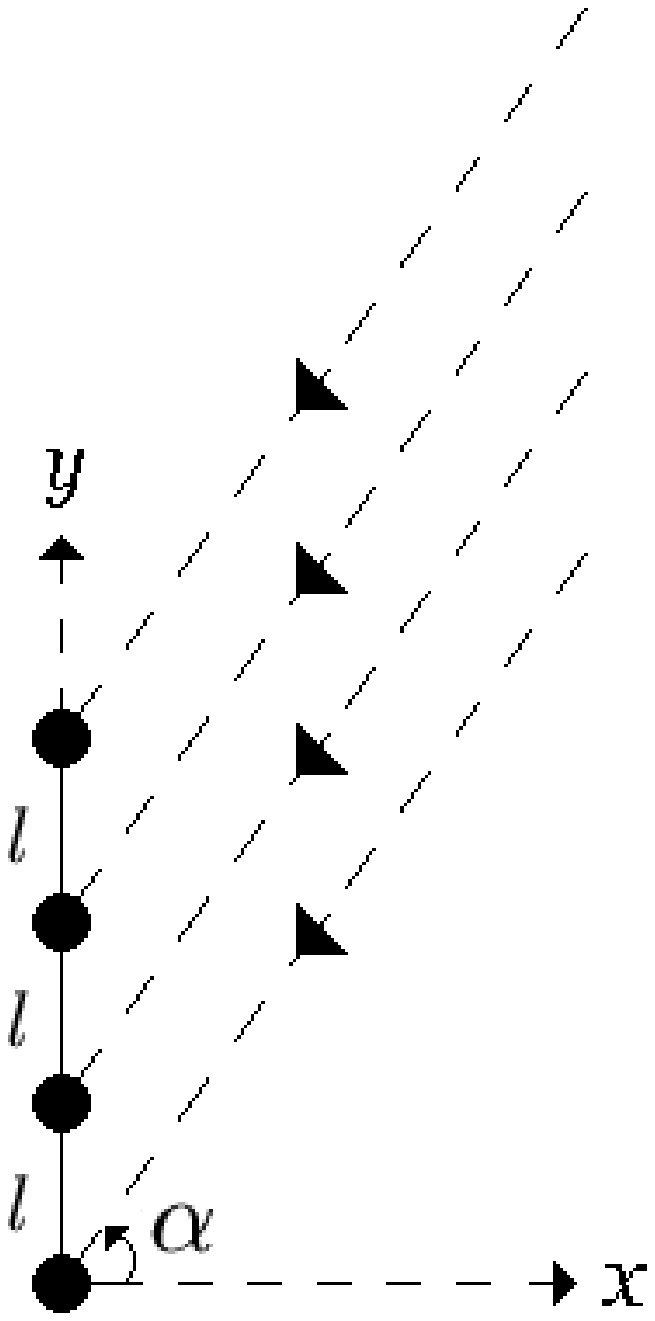}
\caption{A ULA configuration and a signal arriving at the ULA with
angle $\alpha$.} \label{fig:ULA}
\end{center}
\end{figure}

Another position related parameter is angle-of-arrival (AOA), which
specifies the angle between two nodes as shown in Fig.
\ref{fig:AOA}.
Commonly, multiple antennas in the form of an antenna array are
employed at a node in order to estimate the AOA of the signal
arriving at that node. The main idea behind AOA estimation via
antenna arrays is that differences in arrival times of an incoming
signal at different antenna elements contain the angle information
for a known array geometry \cite{Sinan_Springer07}. For example, in
a uniform linear array (ULA) configuration, as shown in Fig.
\ref{fig:ULA}, the incoming signal arrives at consecutive array
elements with ${l \sin\alpha}{/c}$ seconds difference\footnote{It is
assumed that the distance between the transmitting and receiving
nodes are sufficiently large so that the incoming signal can be
modeled as a planar wave-front as shown in Fig. \ref{fig:ULA}.},
where $l$ is the inter-element spacing, $\alpha$ is the AOA and $c$
represents the speed of light \cite{ourBook}. Hence, estimation of the time
differences of arrivals provides the angle information.

Since time delay in a narrowband signal can be approximately
represented by a phase shift, the combinations of the phase shifted
versions of received signals at array elements can be tested for
various angles in order to estimate the direction of signal arrival
\cite{Caffery_Book_2000} in a narrowband system. However, for UWB
systems, time delayed versions of received signals should be
considered, because a time delay cannot be represented by a unique
phase value for a UWB signal.

Similar to the RSS case, the theoretical lower bounds on the error
variances of AOA estimates can be investigated in order to determine
accuracy of AOA estimation. The CRLB for the variance of an unbiased
AOA estimate $\hat{\alpha}$ for a ULA with $N_{\rm{a}}$ elements can
be expressed as\footnote{It is assumed that the signal arrives at
each antenna element via a single path. Please refer to
\cite{Esref_AOA} for CRLBs for AOA estimation in multipath
channels.} \cite{Esref_AOA}
\begin{gather}\label{eq:CRLB_AOA3}
\sqrt{{\rm{Var}}\{\hat{\alpha}\}}\geq \frac{\sqrt{3}\,c}
{\sqrt{2}\,\pi\sqrt{\rm{SNR}}\,\beta\sqrt{N_{\rm{a}}(N_{\rm{a}}^2-1)}\,\,l\,\cos\alpha},
\end{gather}
where $\alpha$ is the AOA, $c$ is the speed of light, SNR is the
signal-to-noise ratio for each element\footnote{The same SNR is
assumed for all antenna elements.}, $l$ is the inter-element
spacing, and $\beta$ is the effective bandwidth.

It is observed from (\ref{eq:CRLB_AOA3}) that the accuracy of AOA
estimation increases, as SNR, effective bandwidth, the number of
antenna elements and/or inter-element spacing are increased. It is
important to note that unlike RSS estimates, the accuracy of an AOA
estimate increases linearly with the effective bandwidth, which
implies that UWB signals can facilitate high-precision AOA
estimation.

\subsubsection{Time of Arrival}\label{sec:TOA}

The time of arrival (TOA) of a signal traveling from one node to
another can be used to estimate the distance between those two
nodes. In order to obtain an unambiguous TOA estimate, the nodes
must either have a common clock, or exchange timing information by
certain protocols such as a two-way ranging protocol
\cite{Scholtz2002}, \cite{lindsey}, \cite{IEEE802154aD4}.

The conventional TOA estimation technique involves the use of
correlator or matched filter (MF) receivers \cite{Turin_1960}. In
order to illustrate the basic principle behind these receivers,
consider a scenario in which $s(t)$ is transmitted from a node to
another, and the received signal is expressed as
\begin{gather}\label{eq:recSig_TOA}
r(t)=s(t-\tau)+n(t)~,
\end{gather}
where $\tau$ is the TOA and $n(t)$ is the background noise, which is
commonly modeled as a zero-mean white Gaussian process. A correlator
receiver correlates the received signal $r(t)$ with a local template
$s(t-\hat{\tau})$ for various delays $\hat{\tau}$, and calculates
the delay corresponding to the correlation peak; that is,
\begin{gather}\label{eq:cor_out}
\hat{\tau}_{\rm{TOA}}={\textrm{arg}}\,\underset{\hat{\tau}}{\max}
\int r(t)\,s\left(t-\hat{\tau}\right){\rm{d}}t~.
\end{gather}
It is clear from (\ref{eq:recSig_TOA}) and (\ref{eq:cor_out}) that
the correlator output is maximized at $\hat{\tau}=\tau$ in the
absence of noise. However, the presence of noise can result in
erroneous TOA estimates.

\begin{figure}
\begin{center}
\includegraphics[width=0.75\textwidth]{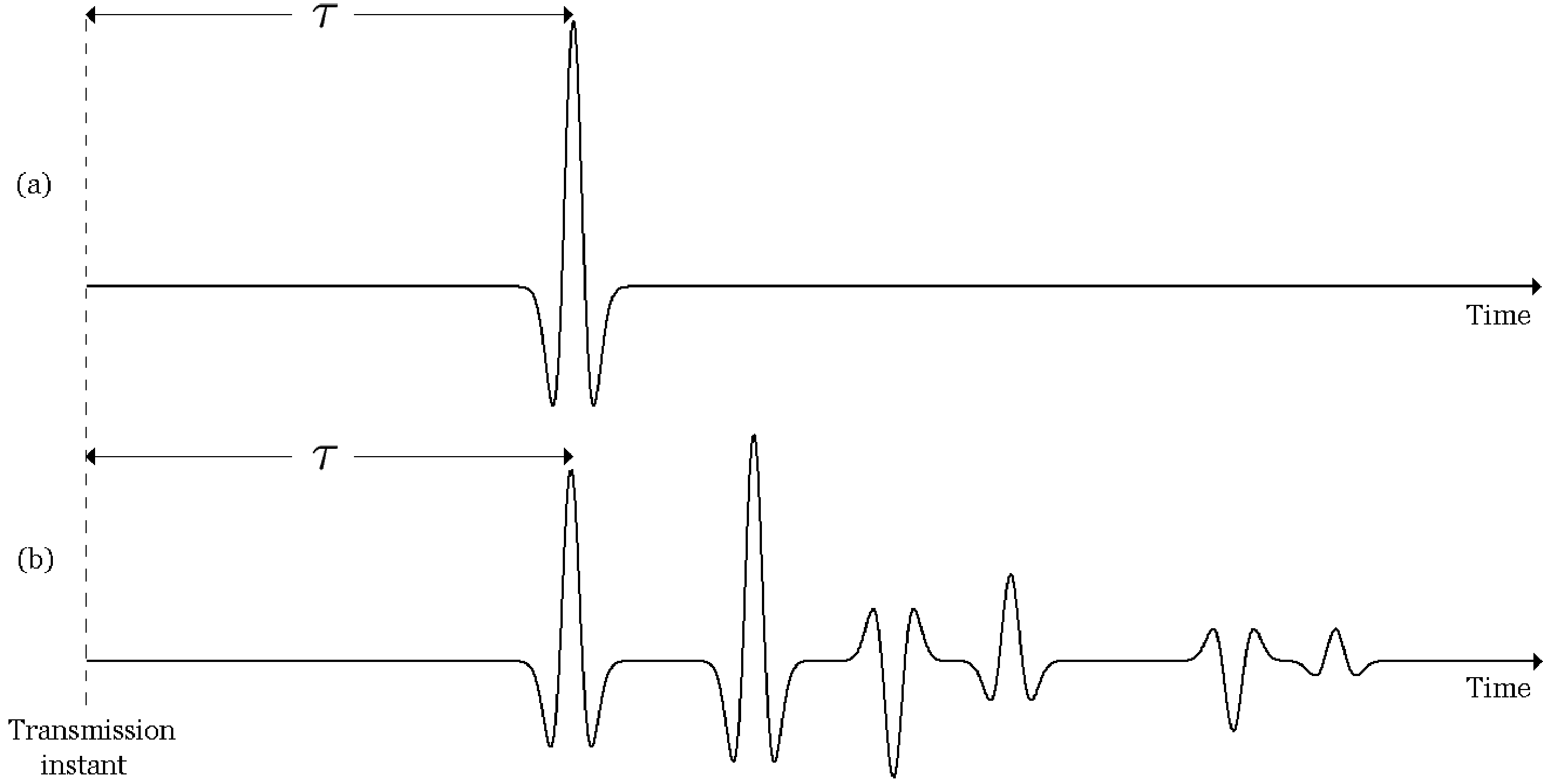}
\caption{a) Received signal in a single-path channel. b) Received
signal over a multipath channel. Noise is not shown in the figure.}
\label{fig:MP}
\end{center}
\end{figure}

Similar to the correlator receiver, the MF receiver employs a filter
that is matched to the transmitted signal and estimates the instant
at which the filter output attains its largest value, which results
in (\ref{eq:cor_out}) as well. Both the correlator and the MF
approaches are optimal\footnote{In the sense that they achieve the
CRLB for TOA estimation asymptotically for large SNRs and/or
effective bandwidths \cite{Yihong_TVT_2006}.} for the signal model
in (\ref{eq:recSig_TOA}) (Fig. \ref{fig:MP}-(a)). However, in
practical systems, the signal arrives at the receiver via multiple
signal paths, as shown in Fig. \ref{fig:MP}-(b).
In those cases, the optimal template signal for a correlator
receiver (or the optimal impulse response for an MF receiver) should
include the overall effects of the channel; that is, it should be
equal to the received signal (with no noise) that consists of all
incoming signal paths. Since the parameters of the multipath channel
are not known at the time of TOA estimation, the conventional
schemes use the transmitted signal as the template, which makes them
suboptimal in general. In this case, selection of the correlation
peak as in (\ref{eq:cor_out}) can result in significant errors, as
the first signal path may not be the strongest signal one, as shown
in Fig. \ref{fig:MP}-(b). In order to achieve accurate TOA
estimation in multipath environments, first-path detection
algorithms are proposed for UWB systems \cite{Scholtz2002},
\cite{Guvenc_ICU_05}-\nocite{Gezici2005}\cite{Yang_Tcomm_2006},
which try to select the first incoming signal path instead of the
strongest one.

Accuracy limits for TOA estimation can be quantified by the CRLB,
which is given by the following\footnote{The CRLBs for TOA
estimation in multipath channels are studied in
\cite{Botteron_2004}, \cite{Gezici_SPM_2005}.} for the signal model
in (\ref{eq:recSig_TOA}):
\begin{gather}\label{eq:CRLB_TOA}
\sqrt{\textrm{Var}(\hat{\tau})}\geq
\frac{1}{2\sqrt{2}\pi\sqrt{\textrm{SNR}}\,\beta}~,
\end{gather}
where $\hat{\tau}$ represents an unbiased TOA estimate, SNR is the
signal-to-noise ratio, and $\beta$ is the effective bandwidth
\cite{Poor}, \cite{cook}. The CRLB expression in (\ref{eq:CRLB_TOA})
implies that the accuracy of TOA estimation increases with SNR and
effective bandwidth. Therefore, large bandwidths of UWB signals can
facilitate very precise TOA measurements. As an example, for the second derivative of a Gaussian pulse \cite{Ramirez} with a pulse width of $1$ ns, the
CRLB for the standard deviation of an unbiased range estimate
(obtained by multiplying the TOA estimate by the speed of light) is
less than a centimeter at an SNR of $5$ dB.

\subsubsection{Time Difference of Arrival}\label{sec:TDOA}

Another position related parameter is the difference between the
arrival times of two signals traveling between the target node and
two reference nodes. This parameter, called time difference of
arrival (TDOA), can be estimated unambiguously if there is
synchronization among the reference nodes \cite{Caffery_Book_2000}.

One way to estimate TDOA is to obtain TOA estimates related to the
signals traveling between the target node and two reference nodes,
and then to obtain the difference between those two estimates. Since
the reference nodes are synchronized, the TOA estimates contain the
same timing offset (due to the asynchronism between the target node
and the reference nodes). Therefore, the offset terms cancel out as
the TDOA estimate is obtained as the difference between the TOA
estimates \cite{Sinan_Springer07}.

When the TDOA estimates are obtained from the TOA estimates as
described above, the accuracy limits can be deduced from the CRLB
expression in the previous section. Namely, it can be concluded that
the accuracy of TDOA estimation improves as effective bandwidth
and/or SNR increases \cite{Sinan_Springer07}.

Another way to obtain the TDOA parameter is to perform
cross-correlations of the two signals traveling between the target
node and the reference nodes, and to calculate the delay
corresponding to the largest cross-correlation value
\cite{Caffery_TVT_1998}. That is,
\begin{gather}\label{eq:TDOA_est_CC}
\hat{\tau}_{\rm{TDOA}}={\rm{arg}}\,\underset{\tau}\max
\left|\int_{0}^{T}r_1(t)\,r_2(t+\tau){\rm{d}}t\right|~,
\end{gather}
where $r_i(t)$, for $i=1,2$, represents the signal traveling between
the target node and the $i$th reference node, and $T$ is the
observation interval.

\subsubsection{Other Position Related Parameters}

In some positioning systems, a combination of position related
parameters, studied in the previous sections, can be utilized in
order to obtain more information about the position of the target
node. Examples of such \textit{hybrid} schemes include TOA/AOA
\cite{Cong_TWC_2002}, TOA/RSS \cite{zafer_CRLB}, TDOA/AOA
\cite{Cong_TWC_2002}, and TOA/TDOA \cite{Reza_2000_hybrid_thesis}
positioning systems.

In addition to the algorithms that estimate RSS, AOA and T(D)OA
parameters or their combinations, another scheme for position
related parameter estimation involves measurement of multipath power
delay profile (PDP)\footnote{Similar to the PDP parameter, the
multipath angular power profile parameter can be estimated at nodes
with antenna arrays.} or channel impulse response (CIR) related to a
received signal
\cite{Nerguizian_2001}-\nocite{Triki_2006,Althaus_VTC_2005}\cite{Nerguizian_TWC_2006}.
In certain cases, PDP or CIR parameters can provide significantly
more information about the position of the target node than the
previously studied schemes \cite{Sinan_Springer07}. For example, a
single TOA measurement provides information about the distance
between a target and a reference node, which determines the position
of the target on a circle; however, CIR information can directly
determine the position of the target node in certain cases if the
observed channel profile is unique for the given environment. In
order to obtain position estimates from CIR (or PDP) parameters, a
database consisting of previous PDP (or CIR) measurements at a
number of known positions are commonly required. In addition,
estimation of PDP/CIR information is usually more complex than the
estimation of the previously studied parameters \cite{ourBook}.


\subsection{Position Estimation}\label{sec:posiEst}

As shown in Fig. \ref{fig:direct_OR_2step}-(b), in the second step
of a two-step positioning algorithm, the position of the target node
is estimated based on the position related parameters estimated in
the first step. Depending on the presence of a database (training
data), two types of position estimation schemes can be considered
\cite{Sinan_Springer07}:
\begin{itemize}
\item \textit{Geometric} and \textit{statistical} techniques estimate
the position of the target node from the signal parameters,
estimated in the first step of the positioning algorithm, via
geometric relationships and statistical approaches, respectively.
\item \textit{Mapping (fingerprinting)} techniques employ a
database, which consists of previously estimated signal parameters
at known positions, to estimate the position of the target node.
Commonly, the database is obtained beforehand by a training
(off-line) phase.
\end{itemize}

\subsubsection{Geometric and Statistical Techniques}\label{sec:Geo_Stat}

\begin{figure}
\begin{center}
\includegraphics[width=0.3\textwidth]{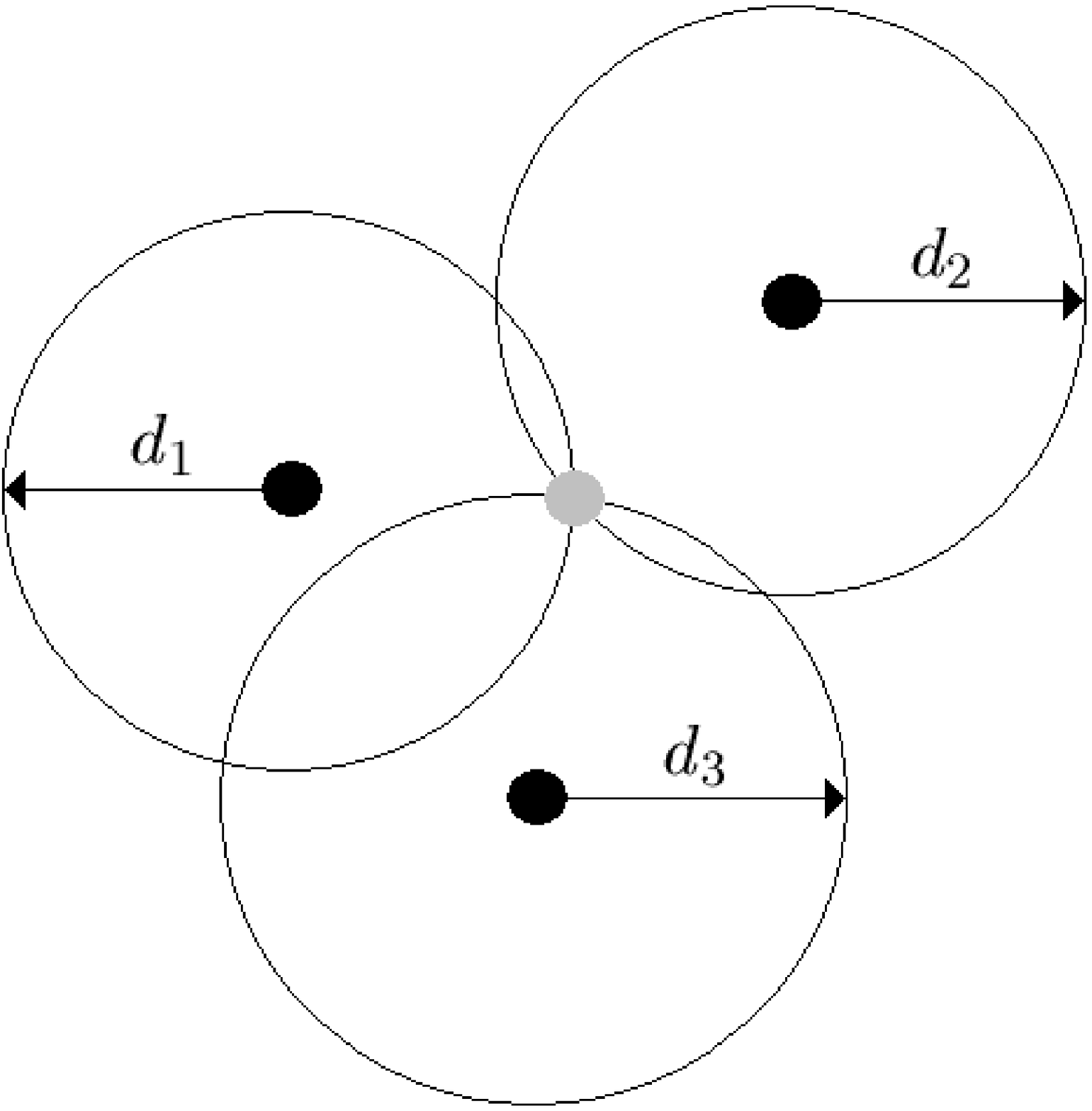}
\caption{Position estimation via trilateration.} \label{fig:trilat}
\end{center}
\end{figure}

\begin{figure}
\begin{center}
\includegraphics[width=0.25\textwidth]{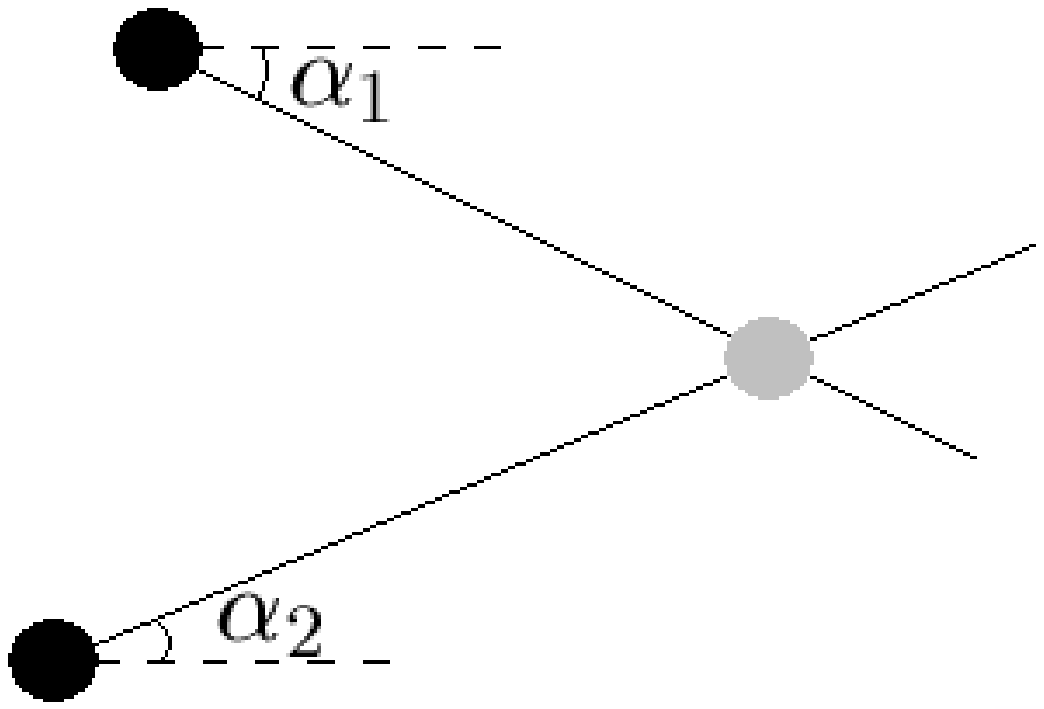}
\caption{Position estimation via triangulation.} \label{fig:triang}
\end{center}
\end{figure}

\begin{figure}
\begin{center}
\includegraphics[width=0.35\textwidth]{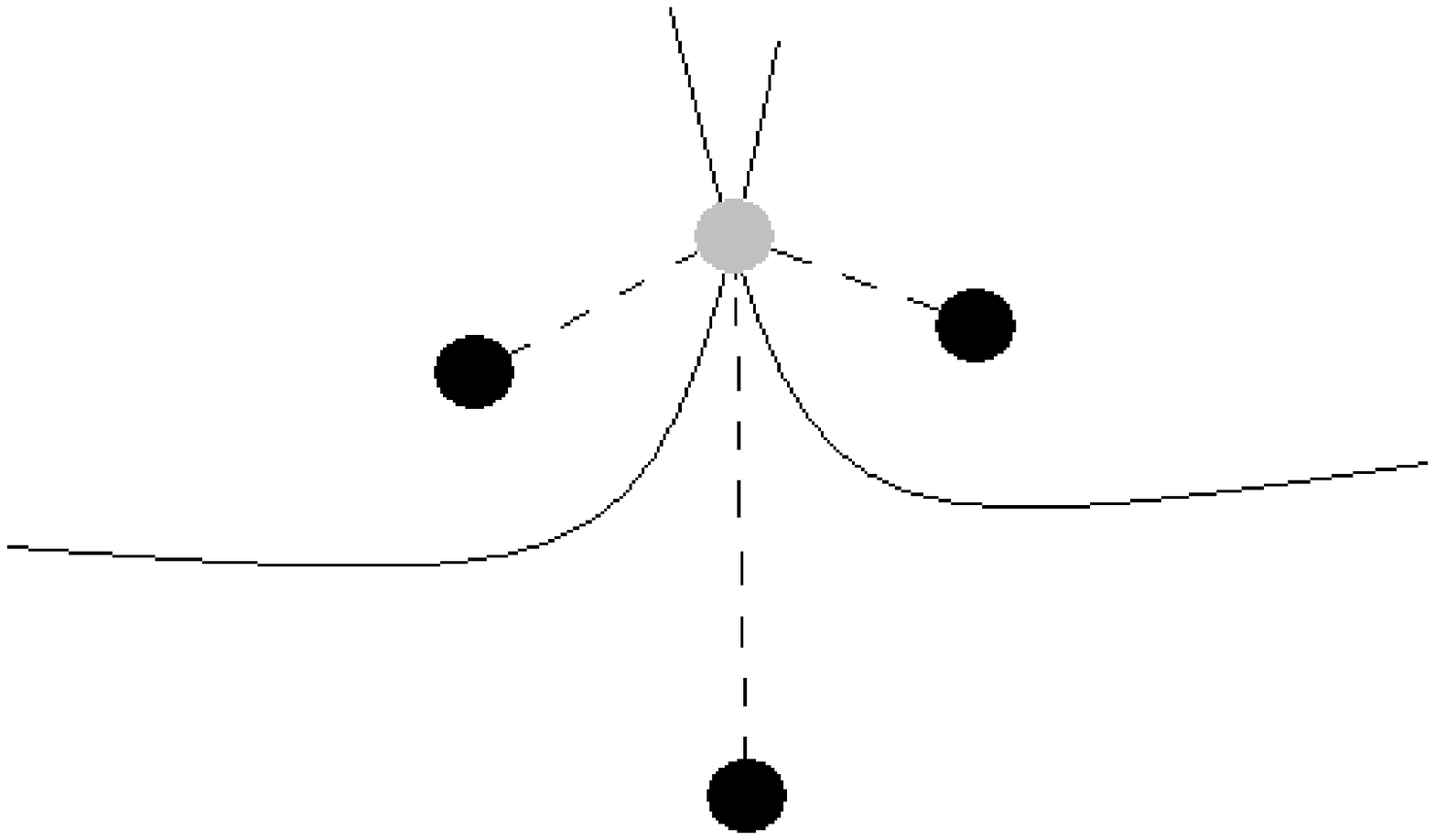}
\caption{Position estimation based on TDOA measurements.}
\label{fig:hyper}
\end{center}
\end{figure}

Geometric techniques for position estimation determine the position
of a target node according to geometric relationships. For example,
a TOA (or an RSS) measurement specifies the range between a
reference node and a target node, which defines a circle for the
possible positions of the target node. Therefore, in the presence of
three measurements, the position of the target node can be
determined by the intersection of three circles\footnote{A
two-dimensional positioning scenario is considered for the
simplicity of illustrations.} via \textit{trilateration}, as shown
in Fig. \ref{fig:trilat}.
On the other hand, two AOA measurements between a target node and
two reference nodes can be used to determine the position of the
target node via \textit{triangulation} (Fig. \ref{fig:triang}).
For TDOA based positioning, each TDOA parameter defines a hyperbola
for the position of the target node. Hence, in the presence of three
reference nodes, two TDOA measurements can be obtained with respect
to one of the reference nodes. Then, the intersection of two
hyperbolas, corresponding to two TDOA measurements, determines the
position of the target node as shown in Fig. \ref{fig:hyper}.
In TDOA based positioning, the position of the target node may not
always be determined uniquely depending on the geometrical
conditioning of the nodes \cite{Sayed_SPM_2005},
\cite{Caffery_Book_2000}.

Geometric techniques can be employed for hybrid positioning systems,
such as TDOA/AOA \cite{Cong_TWC_2002} or TOA/TDOA
\cite{Reza_2000_hybrid_thesis}, as well. For example, if a reference
node obtains both TOA and AOA parameters from a target node, it can
determine the position of the target node as the intersection of a
circle, defined by the TOA parameter, and a straight line, defined
by the AOA parameter \cite{Sinan_Springer07}.

\begin{figure}
\begin{center}
\includegraphics[width=0.3\textwidth]{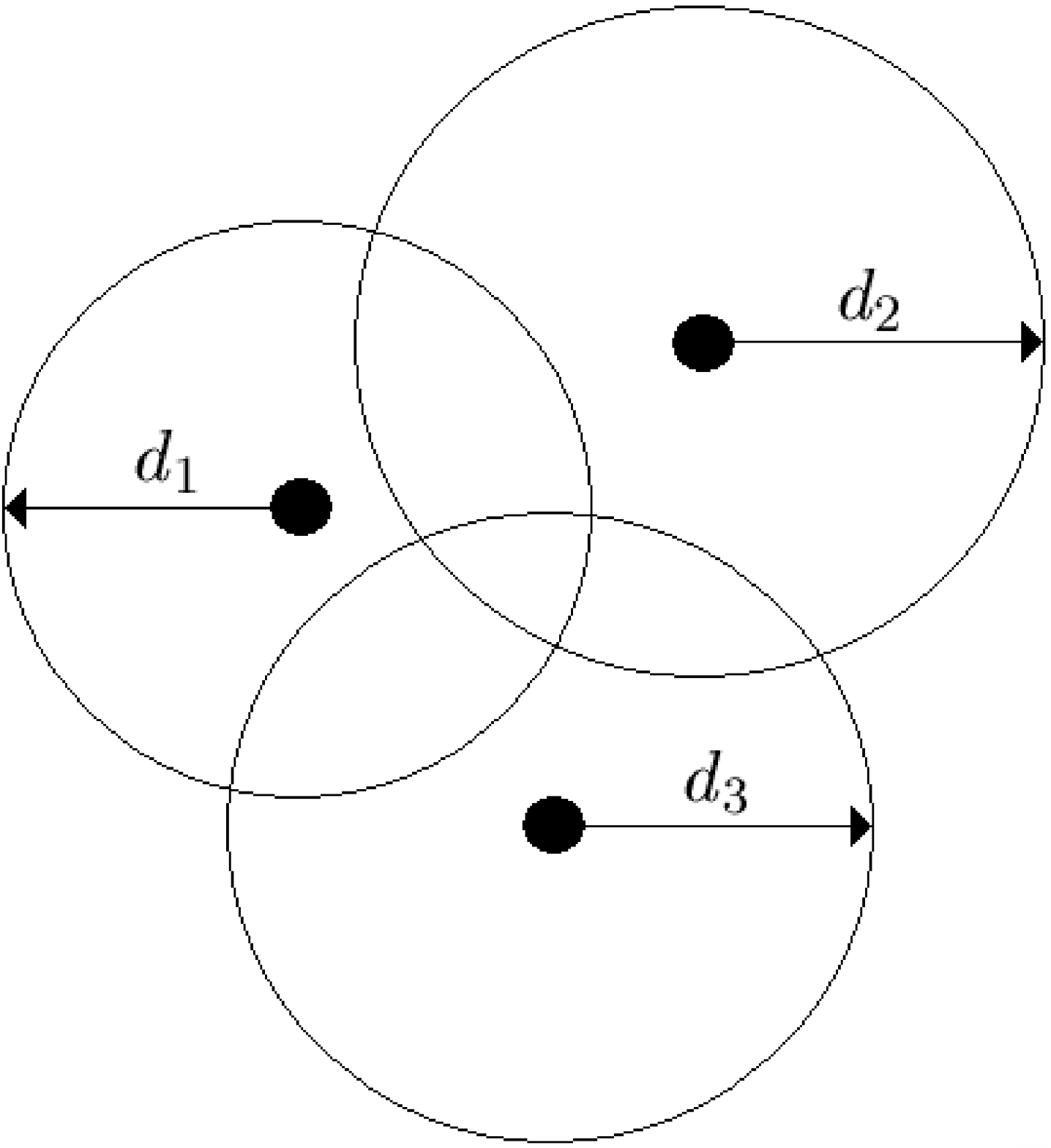}
\caption{Position estimation ambiguities due to multiple
intersections of position lines.} \label{fig:trilat_err}
\end{center}
\end{figure}

Although the geometric techniques provide an intuitive approach for
position estimation in the absence of noise, they do not present a
systematic approach for position estimation based on noisy
measurements. In practice, position related parameter measurements
include noise, which results in the cases that the position lines
intersect at multiple points, instead of a single point, as shown in
Fig. \ref{fig:trilat_err}.
In such cases, the geometric techniques do not provide any insight
as to which point to choose as the position of the target node \cite{ourBook}. In
addition, as the number of reference nodes increases, the
number of intersections can increase even further. 
In other words, the geometric techniques do not provide an efficient
data fusion mechanism; i.e., cannot utilize multiple parameter
estimates in an efficient manner \cite{Sinan_Springer07}.

Unlike the geometric techniques, the statistical techniques present
a theoretical framework for position estimation in the presence of
multiple position related parameter estimates with or without noise.
To formulate this generic framework, consider the following model
for the parameters obtained from the first step of a two-step
positioning algorithm \cite{Sinan_Springer07}:
\begin{gather}\label{eq:measModel}
z_i=f_i(x,y)+\eta_i\,,\quad\quad i=1,\ldots,N_{\rm{m}}\,,
\end{gather}
where $N_{\rm{m}}$ is the number of parameter estimates, $f_i(x,y)$
is the true value of the $i$th signal parameter, which is a function
of the position of the target, $(x,y)$, and $\eta_i$ is the noise at
the $i$th estimation. Note that $N_{\rm{m}}$ is equal to the number
of reference nodes for RSS, AOA and TOA based positioning, whereas
it is one less than the number of reference nodes for TDOA based
positioning since each TDOA parameter is estimated with respect to
one reference node. Depending on the type of the position related
parameter, $f_i(x,y)$ in (\ref{eq:measModel}) can be expressed
as\footnote{Time parameters are converted to distance parameters by
scaling by the speed of light.}
\begin{gather}\label{eq:f_i}
f_i(x,y)=
\begin{cases}
\sqrt{(x-x_i)^2+(y-y_i)^2}, &\textrm{TOA/RSS}\\
\tan^{-1}\left(\frac{y-y_i}{x-x_i}\right), &\textrm{AOA}\\
\sqrt{(x-x_i)^2+(y-y_i)^2}-\sqrt{(x-x_0)^2+(y-y_0)^2},
&\textrm{TDOA}
\end{cases},
\end{gather}
where $(x_i,y_i)$ is the position of the $i$th reference node and
$(x_0,y_0)$ is the reference node, relative to which the TDOA
parameters are estimated.

In vector notations, the model in (\ref{eq:measModel}) can be
expressed as
\begin{gather}\label{eq:meas_vec}
\zz=\ff(x,y)+\bet,
\end{gather}
where $\zz=[z_1\cdots z_{N_{\rm{m}}}]^T$, $\ff(x,y)=[f_1(x,y)\cdots
f_{N_{\rm{m}}}(x,y)]^T$ and
$\bet=[\eta_1\cdots\eta_{N_{\rm{m}}}]^T$.

A statistical approach estimates the most likely position of the
target node based on the reliability of each parameter estimate,
which is determined by the characteristics of the noise corrupting
that estimate. Depending on the amount of information on the noise
term $\bet$ in (\ref{eq:meas_vec}), the statistical techniques can
be classified as \textit{parametric} and \textit{nonparametric}
techniques \cite{ourBook}. For the parametric techniques, the probability density
function (PDF) of the noise $\bet$ is known except for a set of
parameters, denoted by $\blam$. However, for the nonparametric
techniques, there is no information about the form of the noise PDF.
Although the form of the PDF is unknown in the nonparametric case,
there can still be some generic information about some of its
parameters \cite{Cong_2005_NLOS}, such as its variance and symmetry
properties, which can be employed for designing nonparametric
estimation rules, such as the least median of squares technique in
\cite{Casas_2006_Robust}, the residual weighting algorithm in
\cite{Chen_WCNC_1999} and the variance weighted least squares
technique in \cite{Caffery_CM_1998}. In addition, mapping techniques
(to be studied in Section \ref{sec:Mapping}), such as
$k$-nearest-neighbor ($k$-NN) estimation, support vector regression
(SVR), and neural networks, are also nonparametric as they estimate
the position based on a training database without assuming a
specific form for the noise PDF.

Considering the parametric approaches, let the vector of unknown
parameters be represented by $\bthe$, which consists of the position
of the target node, as well as the unknown parameters of the noise
distribution\footnote{In general, the noise components may also
depend on the position of the target node, in which case $\bthe$
includes the union of $x$, $y$ and the elements in $\blam$.}; i.e.,
$\bthe=\left[x\,\,y\,\,\blam^T\right]^T$. Depending on the
availability of prior information on $\bthe$, Bayesian or maximum
likelihood (ML) estimation techniques can be applied
\cite{Caffery_2002_ML_Bayesian}.

For the Bayesian approach, there exists \textit{a priori}
information on $\bthe$, represented by a prior probability
distribution $\pi(\bthe)$. A Bayesian estimator obtains an estimate
of $\bthe$ by minimizing a specific cost function \cite{Poor}. Two
common Bayesian estimators are the minimum mean square error (MMSE)
and the maximum \textit{a posteriori} (MAP)
estimators\footnote{Although MAP estimation is not properly a
Bayesian approach, it still fits within the Bayesian framework
\cite{Poor}.}, which estimate $\bthe$, respectively, as
\begin{align}\label{eq:MMSE}
&\hat{\bthe}_{\rm{MMSE}}={\rm{E}}\left\{\bthe|\,\zz\right\},\\\label{eq:MAP}
&\hat{\bthe}_{\rm{MAP}}=\arg\underset{\bthe}\max\,p(\zz|\bthe)\pi(\bthe)\,,
\end{align}
where ${\rm{E}}\left\{\bthe|\,\zz\right\}$ is the conditional
expectation of $\bthe$ given $\zz$, and $p(\zz|\bthe)$ represents
the conditional PDF of $\zz$ given $\bthe$ \cite{Sinan_Springer07}.

For the ML approach, there is no prior information on $\bthe$. In
this case, an ML estimator calculates the value of $\bthe$ that
maximizes the likelihood function $p(\zz|\bthe)$; i.e.,
\begin{gather}\label{eq:ML}
\hat{\bthe}_{\rm{ML}}=\arg\underset{\bthe}\max\,p(\zz|\bthe)\,.
\end{gather}
Since $\ff(x,y)$ is a deterministic function, the likelihood
function can be expressed as
\begin{gather}\label{eq:likeli_1}
p(\zz|\bthe)=p_\bet(\zz-\ff(x,y)\,|\,\bthe)\,,
\end{gather}
where $p_\bet(\cdot\,|\,\bthe)$ denotes the conditional PDF of the
noise vector given $\bthe$.

In statistical approaches, the exact form of the position estimator
depends on the noise statistics. An example is studied below \cite{ourBook}.

\textbf{Example 1} Assume that the noise components are independent.
Then, the likelihood function in (\ref{eq:likeli_1}) can be
expressed as
\begin{gather}\label{eq:likelihood_ind}
p(\zz|\bthe)=\prod_{i=1}^{N_{\rm{m}}}p_{\eta_i}(z_i-f_i(x,y)\,|\,\bthe)\,,
\end{gather}
where $p_{\eta_i}(\cdot\,|\,\bthe)$ represents the conditional PDF
of the $i$th noise component given $\bthe$. In addition, if the
noise PDFs are given by zero mean Gaussian random variables,
\begin{gather}\label{eq:gausNoise}
p_{\eta_i}(n)=\frac{1}{\sqrt{2\pi}\,\sigma_i}\exp\left(-\frac{n^2}{2\sigma_i^2}\right)~,
\end{gather}
for $i=1,\ldots,N_{\rm{m}}$, with known variances, the likelihood
function in (\ref{eq:likelihood_ind}) becomes
\begin{gather}\label{eq:likelihood_ind_Gaus}
p(\zz|\bthe)=\frac{1}{(2\pi)^{N_{\rm{m}}/2}\prod_{i=1}^{N_{\rm{m}}}\sigma_i}
\exp\left(-\sum_{i=1}^{N_{\rm{m}}}\frac{(z_i-f_i(x,y))^2}{2\sigma_i^2}\right)\,,
\end{gather}
where the unknown parameter vector $\bthe$ is given by
$\bthe=[x\,\,y]^T$. In the absence of any prior information on
$\bthe$, the ML approach can be followed, and the ML estimator in
(\ref{eq:ML}) can obtained from (\ref{eq:likelihood_ind_Gaus}) as
\begin{gather}\label{eq:NLS}
\hat{\bthe}_{\rm{ML}}=\arg\underset{[x\,\,y]^T}\min\,
\sum_{i=1}^{N_{\rm{m}}}\frac{(z_i-f_i(x,y))^2}{\sigma_i^2}~,
\end{gather}
which is the well-known non-linear least-squares (NLS) estimator
\cite{Caffery_Book_2000}, \cite{Sinan_Springer07}. Note that the
terms in the summation are weighted inversely proportional to the
noise variances, as a larger variance means a less reliable
estimate. Among common techniques for solving (\ref{eq:NLS}) are
gradient descent algorithms and linearization techniques via the
Taylor series expansion \cite{Caffery_Book_2000},
\cite{Kim_TVT_2006}. $\square$

In Example 1, the noise components related to different estimates
are assumed to be independent. This assumption is usually valid for
TOA, RSS and AOA estimation. However, for TDOA estimation, the noise
components are correlated, since all TDOA parameters are obtained
with respect to the same reference node. Therefore, TDOA based
systems should be studied through the generic expression in
(\ref{eq:likeli_1}). In addition, the Gaussian model for the noise
terms is not always very accurate, especially for scenarios in which
there is no direct propagation path between the target node and the
reference node. For such NLOS situations, position estimation can be
quite challenging, as will be discussed in Section \ref{sec:NLOS}.

\subsubsection{Mapping Techniques}\label{sec:Mapping}

A mapping technique utilizes a training data set to determine a
position estimation rule (pattern matching algorithm/regression
function), and then uses that rule in order to estimate the position
of a target node for a given set of position related parameter
estimates. Common mapping techniques include $k$-NN, SVR and neural
networks \cite{Nerguizian_TWC_2006},
\cite{McGuire}-\nocite{sinan_SVR,Lin_2005,Kwon_VTC_2004}\cite{Duda_Hart}.

Consider a training data set given by
\begin{gather}\label{eq:train}
{\mathcal{T}}=
\left\{(\mm_1,\llv_1),(\mm_2,\llv_2),...(\mm_{N_{\rm{T}}},\llv_{N_{\rm{T}}})\right\},
\end{gather}
where $\mm_i$ represents the vector of estimated parameters
(measurements) for the $i$th position, $\llv_i$ is the position
(location) vector for the $i$th training data, which is given by
$\llv_i=[x_i\,\,y_i]^T$ for two-dimensional positioning, and
$N_{\rm{T}}$ is the total number of elements in the training set
\cite{Sinan_Springer07}. Depending on the type of the position
related parameters employed in the system, $\mm_i$ can, for example,
consist of RSS parameters measured at the reference nodes when the
target node is at location $\llv_i$. A mapping technique determines
a position estimation rule based on the training set in
(\ref{eq:train}) and then estimates the position of a target node by
using that estimation rule with the measurements related to the
target node.

In order to provide intuition on mapping techniques for position
estimation, the $k$-NN approach can be considered. Let $\llv$ denote
the position of a target node and $\mm$ the measurements (parameter
estimates) related to that node. The $k$-NN scheme estimates the
position of the target node according to the $k$ parameter vectors
in $\mathcal{T}$ that have the smallest Euclidian distances to the
given parameter vector $\mm$. The position estimate $\hat{\llv}$ is
calculated as the weighted sum of the positions corresponding to
those nearest parameter vectors; i.e.,
\begin{gather}\label{eq:kNN}
\hat{\llv}=\sum_{i=1}^{k}{w_i(\mm)\,\llv^{(i)}},
\end{gather}
where $\llv^{(1)},\ldots,\llv^{(k)}$ are the positions corresponding
to the $k$ nearest parameter vectors, $\mm^{(1)},\ldots,\mm^{(k)}$,
to $\mm$, and $w_1(\mm),\ldots,w_k(\mm)$ are the weighting factors
for each position. In general, the weighting factors are determined
according to the parameter vector $\mm$ and the training parameter
vectors $\mm^{(1)},\ldots,\mm^{(k)}$ \cite{McGuire}.

SVR and neural network approaches can also be
considered in the same framework as the $k$-NN technique
\cite{Duda_Hart}. For example, SVR estimates the position also based
on a weighted sum of the positions in the training set. However, the
weights are chosen in order to minimize a risk function that is a
combination of empirical error and regressor complexity.
Minimization of the empirical error corresponds to fitting to the
data in the training set as well as possible, while a constraint on
the regressor complexity prevents the overfitting problem
\cite{SVR_Smola}. In other words, the SVR technique considers the
tradeoff between the empirical error and the generalization
error\footnote{Very complex regressors fit the
training data very closely and therefore may not fit to new
measurements very well, especially for small training data sets.
This is called the generalization (overfitting) problem.}.

In addition to $k$-NN and SVR, neural networks can
also be employed in position estimation problems
\cite{Nerguizian_TWC_2006}, \cite{NN_UWB}. In \cite{NN_UWB}, a UWB
mapping technique based on neural networks is proposed in order to
provide accurate position estimation in mines. In general, in
challenging environments, like mines, measurement models become less
reliable; hence position estimation based on statistical approaches
can result in large errors. Therefore, the mapping techniques can be
preferred in the absence of reliable signal modeling. They can
provide accurate position estimation in environments with
significant multipath and NLOS propagation.

The main limitation of the mapping techniques
compared to statistical and geometric ones is that the training data
set should be large enough and representative of the current
environment. In other words, the training set should be updated at
sufficient frequency, which can be very costly in dynamic
environments. Therefore, mapping techniques are not commonly
employed in outdoor positioning scenarios.

In terms of accuracy, performance of the mapping
techniques compared to the geometric and the statistical techniques
depends on the environment and system parameters. The most important
parameters in a mapping technique are the size and the
representativeness of the training set, and the accuracy of the
regression technique. On the other hand, geometric and statistical
techniques require accurate signal (measurement) models in order to
provide accurate position estimates.


\section{Time-Based Ranging}\label{sec:ranging}

In a two-step positioning algorithm, the positioning accuracy
increases as the position related parameters in the first step are
estimated more precisely. As studied in Section \ref{sec:parEst},
high time resolution of UWB signals can facilitate precise T(D)OA or
AOA estimation; however, RSS estimation provides very coarse range
estimates as observed in Section \ref{sec:RSS}. In addition, AOA
estimation commonly requires multiple antenna elements and increases
the complexity of a UWB receiver. Therefore, timing related
parameters, especially TOA, are commonly preferred for UWB
positioning systems.

In this section, TOA estimation is studied in detail. First, main
sources of errors in TOA estimation are investigated for practical
UWB positioning systems, and then common TOA estimation techniques
are reviewed. As TOA information can be used to obtain the distance,
commonly called ``range'', between two nodes, TOA estimation and
range estimation (or ranging) will be used interchangeably in the
remainder of the paper.

\subsection{Main Sources of Errors}

In a single-path propagation environment with no interfering signals
and no obstructions in between the nodes, extremely accurate TOA
estimation can be performed. However, in practical environments,
signals arrive at a receiver via multiple signal paths, and there
are interfering signals and obstructions in the environment. In
addition, high time resolution of UWB signals, which facilitates
accurate TOA estimation, can cause practical difficulties. Those
error sources for practical TOA estimation are studied in the
following subsections.

\subsubsection{Multipath Propagation}

In a multipath environment, a transmitted signal arrives at the
receiver via multiple signal paths, as shown in Fig. \ref{fig:MP}.
Due to high resolution of UWB signals, pulses received via multiple
paths are usually resolvable at the receiver. However, for
narrowband systems, pulses received via multiple paths overlap with
each other as the pulse duration is considerably larger than the
time delays between the multipath components. This causes a shift in
the delay corresponding to the correlation peak (c.f. eqn.
(\ref{eq:cor_out})) and can result in erroneous TOA estimation. In
order to mitigate those errors, super-resolution time-delay
estimation algorithms, such as that described in \cite{superReso},
were studied for narrowband systems. However, high time resolution
of UWB signals facilitates accurate correlation based TOA estimation
without the use of such complex algorithms. As discussed in Section
\ref{sec:TOA}, first-path detection algorithms can be employed for
UWB systems \cite{Scholtz2002},
\cite{Guvenc_ICU_05}-\nocite{Gezici2005}\cite{Yang_Tcomm_2006} in
order to accurately estimate TOA by determining the delay of the
first incoming signal path.

In order to analyze the effects of multipath
propagation on TOA estimation, accurate characterization of
UWB channels is needed \cite{IEEEChanMod4a},
\cite{PahlavanChannel}-\nocite{Hasari_Sarnoff_06,Alavi_VTC_2005}\cite{Alavi_block_size}. The UWB channel models proposed by the IEEE 802.15.4a channel modeling committee provide statistical information on delays and amplitudes of various signal paths arriving at a UWB receiver \cite{IEEEChanMod4a}. Based on that statistical information or experimental data, TOA estimation errors can be modeled \cite{PahlavanChannel}, \cite{Hasari_Sarnoff_06}. In \cite{PahlavanChannel}, indoor UWB channel measurements are used to propose a statistical model for TOA estimation errors. On the other hand, \cite{Hasari_Sarnoff_06} characterizes the statistical behavior of delay between the first and the strongest signal paths, which is an important parameter for TOA estimation, based on the IEEE 802.15.4a channel models.

\subsubsection{Multiple-Access Interference}

In the presence of multiple users in a given environment, signals
can interfere with each other, and the accuracy of TOA estimation
can degrade. A common way to mitigate the effects of multiple-access
interference (MAI) is to assign different time slots or frequency
bands to different users. However, there can still be interference
among networks that operate at the same time intervals and/or
frequency bands. Therefore, various MAI mitigation techniques, such
as non-linear filtering \cite{Sahinoglu_Eurasip_2006} and training
sequence design \cite{Yang_Tcomm_2006}, are commonly employed.

\subsubsection{Non-Line-of-Sight Propagation}\label{sec:NLOS}

When the direct line-of-sight (LOS) between two nodes is obstructed,
the direct signal component is attenuated significantly such that it
becomes considerably weaker than some other multipath components, or
it cannot even be detected by the receiver
\cite{PahlavanChannel}, \cite{Alavi_VTC_2005}, \cite{PahlavanChannel2}.
For the former case, first-path detection algorithms can still be
utilized in some cases to estimate the TOA accurately. However, in
the latter case, the delay of the first signal path does not
represent the true TOA, as it includes a positive bias, called
non-line-of-sight (NLOS) error. Mitigation of NLOS errors is one of
the most challenging tasks in accurate TOA estimation.

In order to facilitate accurate positioning in NLOS environments,
mapping techniques discussed in Section \ref{sec:Mapping} can be
employed. As the training data set obtained from the environment
implicitly contains information about NLOS propagation, mapping
techniques have a certain degree of robustness against NLOS errors.

In the presence of statistical information about NLOS errors,
various NLOS identification and mitigation algorithms can be
employed \cite{Gezici_SPM_2005},
\cite{PahlavanChannel}. For example, in
\cite{Wylie_UPC_1996}, the observation that the variance of TOA
measurements in the NLOS case is usually considerably larger than
that in the LOS case is used in order to identify NLOS situations,
and then a simple LOS reconstruction algorithm is employed to reduce
the positioning error. In addition, statistical techniques are
studied in \cite{Mandayam_VTC_1998} and \cite{Gezici_VTC_2003} in
order to classify a set of measurements as LOS or NLOS. Finally,
based on various scattering models for a given environment, the
statistics of TOA measurements can be obtained, and then well-known
techniques, such as MAP and ML, can be employed to mitigate the
effects of NLOS errors \cite{Caffery_2002_ML_Bayesian},
\cite{Caffery_2002_ScatteringNLOS}.

\subsubsection{High Time Resolution of UWB Signals}\label{sec:highRes}

Although high time resolution of UWB signals results in very precise
TOA estimation, it also poses certain practical challenges. First,
clock jitter becomes a significant factor that affects the accuracy
of UWB positioning systems \cite{Shimizu}. Due to short durations of
UWB pulses, clock inaccuracies and drifts in target and reference
nodes can affect the TOA estimates.

In addition, high time resolution of UWB signals makes it quite
impractical to sample received signals at or above the Nyquist
rate\footnote{The Nyquist rate is equal to twice the highest
frequency component contained within a signal.}, which is typically
on the order of a few GHz. Therefore, TOA estimation schemes should
make use of low rate samples in order to facilitate low power
designs.

Finally, high time resolution of UWB signals results in a TOA
estimation scenario, in which a large number of possible delay
positions need to be searched in order to determine the true TOA.
Therefore, conventional correlation-based approaches that search
this delay space in a serial fashion become impractical for UWB
signals. Therefore, fast TOA estimation algorithms, as will be
discussed in the next sub-section, are required in order to obtain
TOA estimates in reasonable time intervals.

\subsection{Ranging Algorithms}

\begin{figure}
\begin{center}
\includegraphics[width=0.5\textwidth]{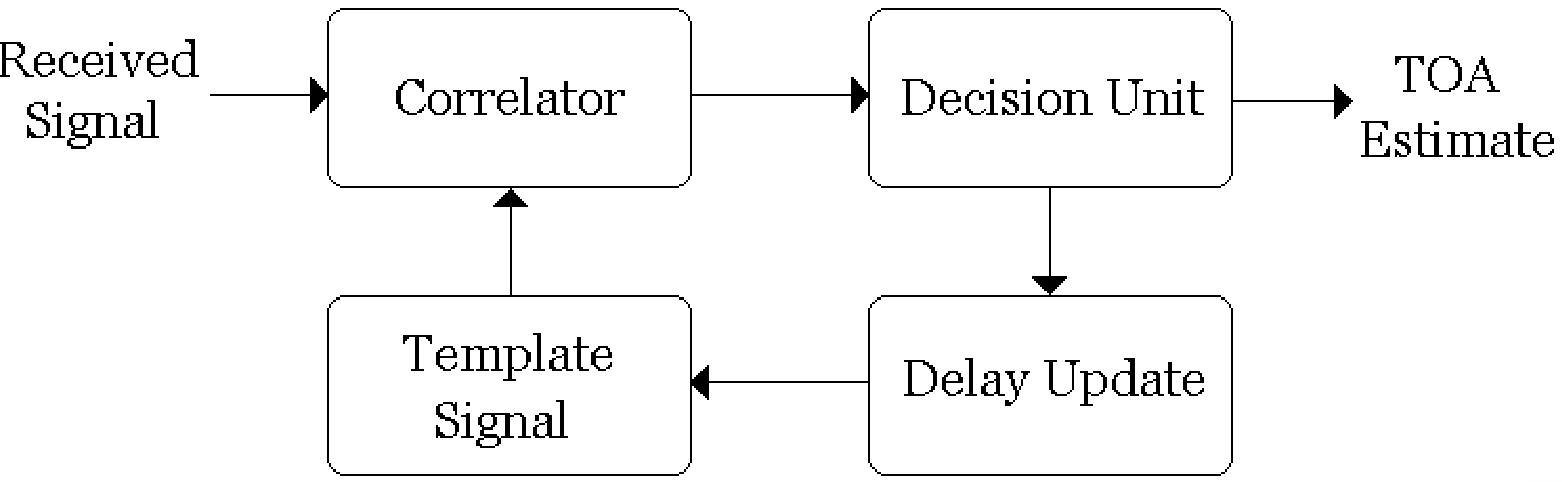}
\caption{A receiver architecture for correlation-based TOA
estimation.} \label{fig:corr}
\end{center}
\end{figure}

As discussed in Section \ref{sec:TOA}, a conventional
correlation-based TOA estimation (equivalently, ranging) algorithm
correlates the received UWB signal with various delayed versions of
a template signal, and obtains the TOA estimate as the delay
corresponding to the correlation peak (Fig. \ref{fig:corr}).
However, in practice, there is a large number of possible signal
delays that need to be searched for the correlation peak due to high
time resolution of UWB signals, and also the correlation peak may
not always correspond to the true TOA. Therefore, a serial search
strategy can be employed, which estimates the delay corresponding to
the first correlation output that exceeds a certain threshold
\cite{Caffery_Book_2000}, \cite{Guvenc_ICU_05},
\cite{Dardari_ICC_2006}.
However, also the serial search approach can take a very long time
to obtain a TOA estimate in many cases \cite{Somayazulu_2002}. In
order to speed up the estimation process, different search
strategies, such as random search or bit reversal search, can be
employed \cite{Homier_UWBST'02}. For example, in a random search
strategy, possible signals delays are selected randomly and tested
for the TOA. In the presence of multipath propagation, the random
search strategy can reduce the time to obtain a rough TOA estimate;
that is, to determine the delay of a signal path (not necessarily
the first one). Then, fine TOA estimation can be performed by
searching backwards in time from the detected signal component
\cite{Guvenc_Sarnoff_2006}.

In general, two-step approaches that estimate a rough TOA in the
first step and then obtain a fine TOA estimate in the second step
can provide significant reduction in the amount of time to perform
ranging \cite{Gezici2005}. Commonly, rough TOA estimation in the
first step can be performed by low-complexity receivers rapidly,
which considerably reduces the possible delay positions that need to
be searched for the fine TOA in the second step. For example, in
\cite{Gezici2005}, a simple energy detector is employed for
determining a rough TOA estimation; i.e., for reducing the TOA
search space. Then, the second step searches for the TOA within a
smaller interval determined by the first step. For the second step,
correlation-based first-path detection schemes \cite{Guvenc_ICU_05},
\cite{Scholtz2002}, or statistical change detection approaches
\cite{Gezici2005} can be employed.

As discussed, in Section \ref{sec:highRes}, TOA estimation based on
low rate sampling, compared to the Nyquist rate sampling, is
desirable for low power implementations. Examples of such low rate
TOA estimators include ones that employ energy detectors
(non-coherent receivers), low rate correlator outputs, or symbol
rate auto-correlation receivers
\cite{Guvenc_UWBNETS_05}-\nocite{Guvenc_ICU_05b,Sahinoglu_Eurasip_2006}\cite{Falsi_Eurasip_2006},
\cite{Yang_Tcomm_2006}.


\section{Practical Considerations}\label{sec:prac}

After studying theoretical aspects and estimation algorithms for UWB
positioning and ranging systems, we now consider practical issues
related to the design of UWB ranging signals, and hardware issues
for UWB transmitters and receivers.

\subsection{Signal Design}\label{sec:SigDesign}

In order to meet certain performance requirements under practical
and regulatory constraints, UWB ranging signals should be designed
appropriately \cite{ourBook}. The main performance criterion in a ranging system is ranging accuracy, which is commonly quantified by the
root-mean-square error (RMSE) given by
\begin{gather}\label{eq:RMSE1}
\textrm{RMSE}=\sqrt{\rmE\left\{(\hat{d}-d)^2\right\}}~,
\end{gather}
where $\hat{d}$ is the range estimate and $d$ is the true range. In
other words, the RMSE is defined as the square root of the average
value of the squared error\footnote{A more generic accuracy metric
is the cumulative distribution function (CDF) of the ranging error,
which specifies the probability that the ranging error is smaller
than a given threshold value for all possible thresholds.}. In
practice, the expected value in (\ref{eq:RMSE1}) is approximated by
the sample mean of the squared error; i.e.,
\begin{gather}\label{eq:RMSE2}
\textrm{RMSE}\approx\sqrt{\frac{1}{N}\sum_{i=1}^{N}
\left(\hat{d}_i-d_i\right)^2}~,
\end{gather}
where $d_i$ and $\hat{d}_i$ are, respectively, the true range and
the range estimate for the $i$th measurement, for $i=1,\ldots,N$.

In addition to ranging accuracy, the duration of a ranging signal is
another important parameter for UWB ranging systems \cite{ourBook}. For small
durations of ranging signals, range estimates can be obtained
quickly and also more signal resources can be allocated for data
transmission if the system is performing both ranging and
communications. Intuitively, as the duration of ranging signal
increases, more accurate range (TOA) estimation can be performed.
This can be observed also from the CRLB expression in
(\ref{eq:CRLB_TOA}). To that end, consider a generic ranging signal
structure, as shown in Fig. \ref{fig:signal}, which is expressed as
\begin{gather}\label{eq:rangSig}
s(t)=\sum_{j=-\infty}^{\infty}a_j\,\omega(t-jT_{\rm{f}})~,
\end{gather}
where $\omega(t)$ is a UWB pulse, $T_{\rm{f}}$ is the frame interval
(or, pulse repetition interval), which is commonly considerably
larger than the pulse width, and $a_j$ is a binary, $\{-1,+1\}$, or
ternary, $\{-1,0,+1\}$, code, which is used for interference
robustness and spectral optimization \cite{Gezici_TSP05},
\cite{Gezici_TCOM07}, \cite{Nakache_2002}. For example, in the IEEE
802.15.4a standard, ternary codes are employed for the
synchronization preamble of each packet, which is used for ranging
purposes \cite{IEEE802154aD4}.

According to the FCC regulations, as shown in Fig. \ref{fig:indoor},
there is a limit on the average PSD of a UWB signal. Depending on
the spectral characteristics of the signal in (\ref{eq:rangSig}),
the maximum amount of average power $\Pmax$ that can be transmitted
by a UWB transmitter can be determined from the FCC limit. Then, the
maximum energy of a pulse in a frame can be calculated as
$\Tf\Pmax$. Therefore, for a ranging system that
employs $\Nf$ UWB pulses, the SNR in the CRLB expression in
(\ref{eq:CRLB_TOA}) becomes directly proportional to $T\Pmax$, where
$T=\Nf\Tf$. Hence, as the duration of the ranging signal increases,
better accuracy can be achieved; i.e., the ranging signal duration
and the lower bound on the error variance of unbiased range
estimators are inversely proportional.

Design of a ranging signal, as in (\ref{eq:rangSig}), also requires
an appropriate selection of the frame interval $\Tf$. As studied
above, the energy of a pulse in a frame is proportional to the frame
interval. Hence, for a given pulse width (or the signal bandwidth),
the peak power increases as the frame interval increases. The peak
power is an important parameter in UWB ranging systems due to
practical limitations of integrated circuits \cite{Lakkis_PAR} and
regulatory constraints \cite{FCC_0248}. Therefore, there exists a
practical upper bound on the frame interval in a given UWB system.
On the other hand, very small frame intervals are not desirable
either, as they can result in interference between pulses in
consecutive frames due to the effects of multipath propagation.
Also, large frame intervals can facilitate low power designs, as
some units in the receiver, such as an analog-to-digital converter
(ADC), can be run only when pulses arrive \cite{Lakkis2}.

After determining the length of the ranging signal $T$, and the
frame interval $\Tf$, another important issue is related to
\textit{pulse coding}, which is implemented by the sequence
$\{a_j\}$ in (\ref{eq:rangSig}). Pulse coding is quite important for
reliable range estimation, as coded pulses in a ranging signal can
provide robustness against multipath and multiple-access
interference (MAI). While autocorrelation properties of a code
determine its robustness against multipath interference, its
cross-correlation properties become effective in mitigating MAI \cite{ourBook}. In addition, code length is an important parameter; since better
correlation properties can be obtained with longer codes, but
shorter codes ease the acquisition process \cite{Gezici_WAMI06},
\cite{Gezici_TSP05}.

\subsection{Hardware Issues}\label{sec:hardware}

After the selection of signal parameters, implementation of a UWB
system requires the design of hardware components for UWB
transmitters and receivers. Due to large bandwidths of UWB signals,
conventional hardware design techniques are not applicable to
certain sections of a UWB transmitter/receiver. In this section,
various issues related to hardware design for UWB systems are
briefly investigated.

\begin{figure}
\begin{center}
\includegraphics[width=0.7\textwidth]{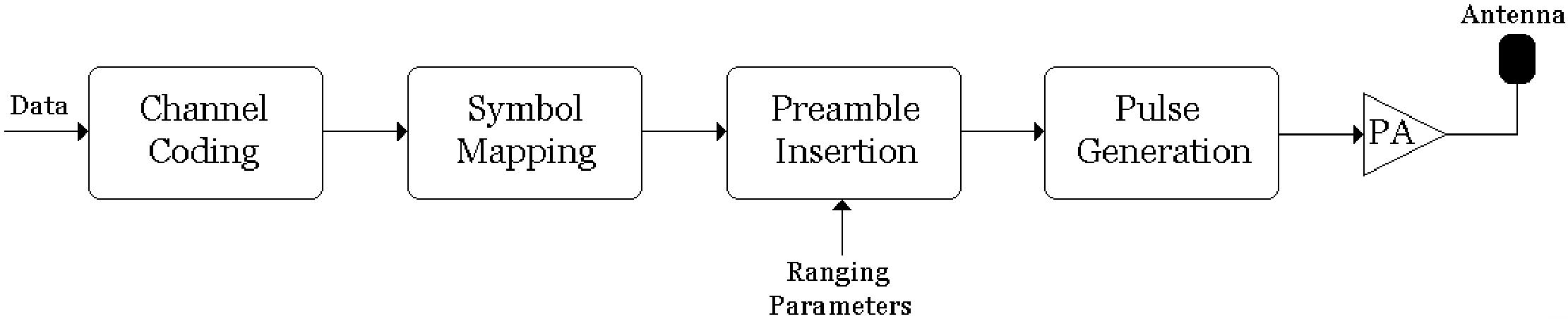}
\caption{Block diagram of a UWB transmitter in a communications
system with ranging capability \cite{ourBook}.} \label{fig:TX}
\end{center}
\end{figure}

UWB systems that provide ranging information commonly perform both
communications and ranging, as in typical IEEE 802.15.4a systems
\cite{IEEE802154aD4}. In other words, such UWB systems provide
low-to-medium rate data communications together with ranging
capability. Fig. \ref{fig:TX} illustrates the block diagram of a UWB
transmitter in such a UWB system \cite{ourBook}.
As shown in the figure, communications data is first coded in order
to provide robustness against the adverse effects of the channel. In
other words, some systematic redundancy is added into the data in
order to recover the correct data at the receiver in the presence of
errors. Then, the coded data is mapped onto specific symbols for
modulation purposes. As an example, the coded data can be mapped
onto binary phase shift keying (BPSK) symbols, which take values
from the set $\{-1,+1\}$. After symbol mapping, ranging related
information is inserted at the beginning of the communications data.
Typically, transmission is performed in terms of \textit{packets},
which contain both communications and ranging signals; i.e., a
certain section of transmission is allocated for ranging signals,
and the remaining is allocated for communications signals. Since
ranging signals commonly constitute the beginning section of each
packet, they are also called \textit{preamble}s\footnote{In a system
that performs both communications and ranging, preamble signals are
used not only for ranging, but also for timing acquisition,
frequency recovery, packet and frame synchronization and channel
estimation.}.

The digital sequence at the output of the preamble insertion block
is converted into an analog UWB pulse sequence by the pulse
generation block. UWB pulse generators can be broadly classified
into two depending on the use of an \textit{up-conversion}
unit\footnote{An up-conversion unit commonly consists of a mixer and
a local oscillator. The incoming signal and the signal generated by
the local oscillator is multiplied using the mixer in order to
perform frequency translation.}. The ones that employ an
up-conversion unit first generate a pulse at baseband, and then
translate the frequency contents of the signal (i.e., ``up-convert''
it) around a desired center frequency
\cite{LO_monoGen}-\nocite{LO_CMOS_DSSS}\cite{LO_PulGen_Gaus_2006}.
On the other hand, some UWB pulse generators can directly generate
the pulses in the desired frequency band without employing any
up-conversion unit. Among such pulse generators are the ones that
generate UWB pulses, such the fifth derivative of a Gaussian pulse,
without any filtering operations \cite{CMOS_Tunable_PulseGen},
\cite{AllDig_PulseGen}, the ones that use antenna for shaping UWB
pulses \cite{ExtLowPowRadio}, \cite{DirectAntMod}, and the ones that
employ filtering for pulse shaping
\cite{PulGen_SingChip_2006}-\nocite{PulGen_CMOSgen,PulGen_BPF,PulGen_BPF2,PulGen_New1}\cite{PulGen_New2}.

After generating UWB pulses, a power amplifier (PA) can be used to
increase the power of the signal delivered to the antenna. For UWB
systems operating under extremely low power regulations, such as the
Japanese regulations for unlicensed use of UWB systems, use of a PA
may not be needed \cite{woPA}. Commonly, PAs can constitute a large
portion of the transmitter power consumption. Hence, it is desirable
to have efficient\footnote{\textit{Efficiency} of a PA is defined as
the ratio between the signal power delivered to the load and the
total power consumed by the amplifier.} PAs in order to minimize the
power consumed at a transmitter
\cite{RF_UWB}-\nocite{distrPA1,feedbackPA1,fullBand_PA}\cite{filterPA2}.

Finally, an antenna unit transmits the UWB signal into space, as
shown in Fig. \ref{fig:TX}. Related to large bandwidths of UWB
signals, UWB antenna design should take a number of issues into
account. First, a UWB antenna should have a wide \textit{impedance
bandwidth}, which is defined as the frequency band over which there
is no more than $10$\% signal loss due to the mismatch between the
transmitter circuitry and the antenna \cite{ourBook}. Ideally, when there is
perfect matching, incoming signal towards the antenna is completely
radiated into space. In order to obtain large impedance bandwidths
for UWB antennas, various bandwidth broadening techniques are
commonly employed. Among those techniques are using specific antenna
geometries such as helix, biconical, and bow-tie structures
\cite{antenTEZ}, beveling or smoothing
\cite{BB_Planar_Antennas}-\nocite{BasebandPulseAnt}\nocite{DiamondAnt}\cite{WidebandAnt},
resistive loading \cite{TXRXuwbAnt}, slotting (or adding a strip)
\cite{Anten_slot1}, \cite{Anten_slot2}, notching, and optimizing
location or structure of the antenna feed
\cite{Anten_feed1}-\nocite{Anten_feed2}\cite{Anten_feed3}. Another
important issue in UWB antenna design is that a UWB antenna should
radiate a pulse that is very similar to the pulse at the feed of the
antenna (or its derivative) so that no significant pulse distortion
occurs \cite{TXRXuwbAnt}. In addition, \textit{radiation
efficiency}, which is defined as the ratio of the radiated power to
the input power at the terminals of the antenna \cite{antenTEZ},
should be quite high so that there is no significant power loss.
Since UWB signals operating under regulatory constraints can
transmit low power signals only, high radiation efficiency of UWB
antennas is needed for ranging/communications at reasonable
distances.

Commonly, planar antennas, such as bow-tie, diamond and square
dipole antennas, and polygonal and elliptical monopole antennas, are
well-suited for UWB systems as they are compact and can be printed
on PCBs (\cite{PlanarAnten} and \cite{StateofArt_ant}, and
references therein). In addition, they can have wide impedance
bandwidths and reasonable pulse distortion if their geometries and
feeding structures are designed in an appropriate fashion
\cite{PlanarAnten}.

\begin{figure}
\begin{center}
\includegraphics[width=0.5\textwidth]{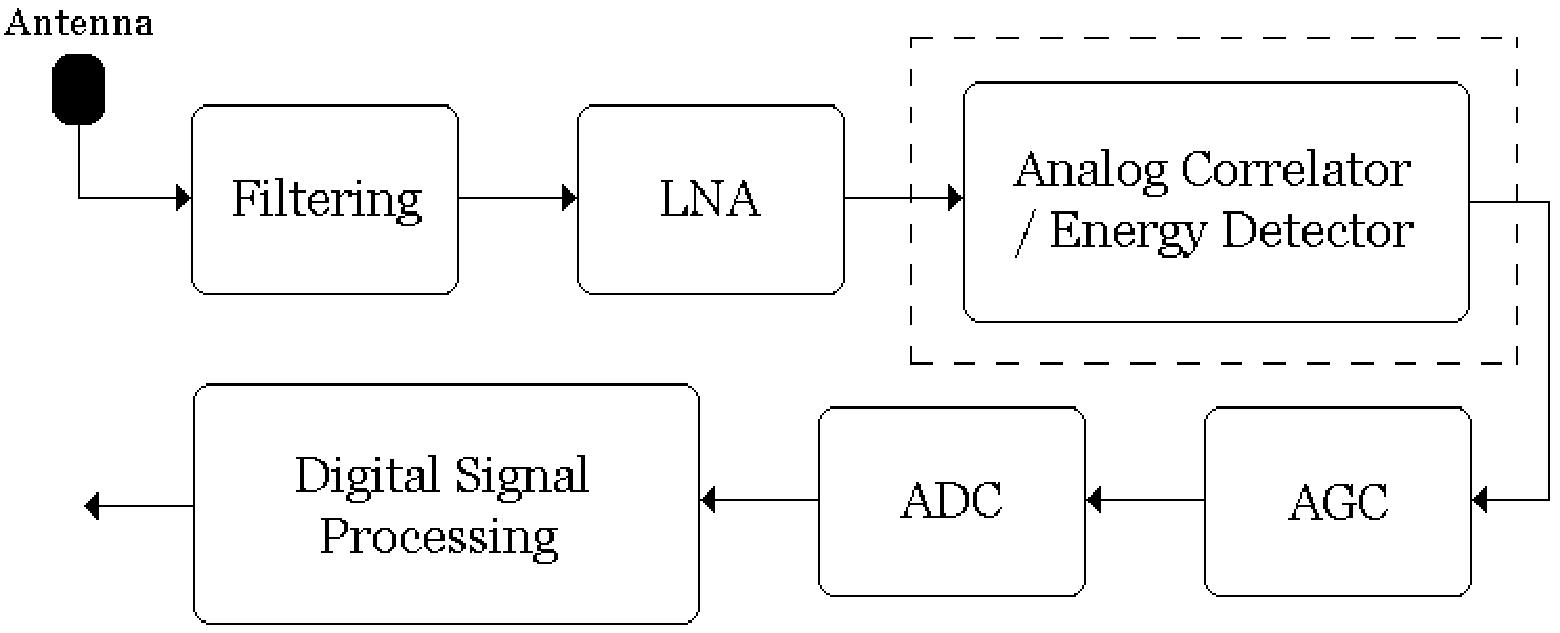}
\caption{Block diagram of a UWB receiver. The unit in the dotted box
exists only when analog correlation or energy detection is to be
performed \cite{ourBook}.} \label{fig:RX}
\end{center}
\end{figure}

Considering the receiver part, UWB signals are collected by a UWB
antenna as shown in Fig. \ref{fig:RX} \cite{ourBook}. Then, the signal is passed
through a band-pass filter (BPF) and a low-noise amplifier (LNA) for
out-of-band noise/interference mitigation and signal amplification,
respectively. At this point, two groups of UWB receivers can be
considered. One group of UWB receivers, called ``all-digital'',
directly convert the analog UWB signal into digital and perform all
the main signal processing operations, such as correlation, in the
digital domain
\cite{channelized_UWB}-\nocite{allDigRX1}\cite{allDigRX2}. In other
words, for all-digital UWB receivers, the units in the dotted box in
Fig. \ref{fig:RX} do no exist. On the other hand, other UWB
receivers perform correlation or energy detection operations
(depending on the receiver type) in the analog domain, and then
perform the conversion from the analog domain to the digital domain
\cite{analogRX1}-\nocite{analogRX2}\cite{analogRX3}. For both
receiver types, analog-to-digital conversion is performed by the ADC
unit, which is preceded by the automatic gain control (AGC) that
adjusts the level of the UWB signal according to ADC specifications.

An ADC obtains samples from the analog signal and quantizes those
samples into a number of levels to represent a digital signal. How
fast those samples are obtained (\textit{sampling rate}), how many
bits are used to represent the digital signal (\textit{resolution}),
and the amount of power dissipation are the main parameters of an
ADC. As the sampling rate and/or the resolution increases, the
complexity and the power dissipation of the ADC increases, as well.
Due to large bandwidths of UWB signals, design of high speed and low
power ADCs is an important issue for UWB receivers.

For UWB receivers that perform correlation (or, energy detection)
operations in the analog domain, ADCs can operate at much lower
rates than the Nyquist rate, as sampling per frame or symbol becomes
sufficient, which faciltates the design of low power UWB receivers
\cite{analogRX1}, \cite{woPA}. However, such receivers commonly
experience performance degradation due to circuit mismatches and
reduced flexibility. For example, the number of correlators are
usually quite limited in the analog implementation, which prevents
the implementation of sophisticated narrowband interference (NBI)
mitigation techniques \cite{channelized_UWB}.

For improved performance, it is desirable to perform
analog-to-digital conversion at an early stage, as in all-digital
UWB receivers. However, for those receiver, very high speed ADCs are
required, as sampling UWB signals at the Nyquist rate requires
obtaining a few billion samples per second (Gsps). Fortunately,
resolution requirement is not as strict as the sampling rate
requirement, and an ADC with a few bits of resolution is usually
sufficient for UWB signals. Specifically, an ADC with more than $4$
bits of resolution provides only marginal improvement over a $4$-bit
ADC for UWB systems \cite{allDigRX1}, \cite{ADCprecision},
\cite{4bitADC_ICU_2005}. In order to meet the fast sampling rate
requirement with the current ADC technology, various channelization
techniques, such as frequency-domain channelization,
\cite{channelized_UWB},
\cite{Monobit_TWC}-\nocite{timeChanADC,orth_ADC}\cite{freqDomainRX},
and subsampling techniques \cite{subsamp1}, \cite{subsamp2} are
commonly employed.

After the ADC, the digital signal samples are
processed in order to estimate a position related parameter, such as
TOA. Then, the position related parameters corresponding to a number of UWB nodes are used to determine the position of the target node. In self-positioning systems, the target node itself calculates the
position, whereas in remote-positioning systems, a central node
calculates the position of the target. In both cases, the complexity
of the position estimation algorithm sets the signal processing
requirements on the related node. For example, if a mapping
technique is implemented, the node needs to manage a training data
set and employ it for position estimation. On the other hand, a
statistical technique does not require training data management, but
may need to solve an optimization problem, such as the NLS algorithm
in (\ref{eq:NLS}).


\section{Conclusion}\label{sec:conc}

In this paper, we have reviewed the problem of position estimation
in UWB wireless systems.   We have considered primarily a two-step
positioning approach, in which the estimation of position related
parameters, such as TOA and AOA, is performed first, followed by
position estimation from those parameters.  We have seen that TOA
systems are particularly well-suited for this purpose, and have
investigated this technique in more depth.  We have also considered
implementation issues for UWB ranging systems.


\bibliographystyle{IEEEtran}
\bibliography{Ranging_procIEEE_3}

\end{document}